\documentclass{article}

 \usepackage[preprint]{neurips_2026}

\usepackage[utf8]{inputenc} 
\usepackage[T1]{fontenc}    
\usepackage{hyperref}       
\usepackage{url}            
\usepackage{booktabs}       
\usepackage{amsfonts}       
\usepackage{nicefrac}       
\usepackage{microtype}      
\usepackage{xcolor}         
\usepackage{placeins}

\usepackage{amsmath}
\usepackage{amsfonts}
\usepackage{amssymb}
\usepackage{wrapfig}
\usepackage{subcaption}
\usepackage{multirow}
 \usepackage{mathtools} 

\usepackage{verbatim}

\usepackage{anyfontsize}
\usepackage{soul}

\usepackage{microtype}
\usepackage{graphicx}
\usepackage{booktabs} 
\usepackage{multirow}
\usepackage{amsmath,amssymb}
\usepackage{booktabs}
\usepackage{caption,subcaption}
\usepackage{array}  

\usepackage{wrapfig}
\usepackage{xcolor}
\definecolor{mygreen}{HTML}{3cb44b}
\definecolor{skyblue}{HTML}{beffff}
\definecolor{lightgreen}{HTML}{90ee90}

\usepackage{color, colortbl}

\definecolor{emerald}{rgb}{0.31, 0.78, 0.37}

\usepackage{tcolorbox}
\usepackage{enumitem}
\setitemize{itemsep=10pt,topsep=0pt,parsep=0pt,partopsep=0pt}
\usepackage{colortbl}

\usepackage{xcolor}
\definecolor{mygreen}{HTML}{3cb44b}
\colorlet{myyellow}{green!10!orange!90!}
\makeatletter

\usepackage{tikz}
\usetikzlibrary{arrows,shapes,snakes,automata,backgrounds,fit,petri}
\usepackage{adjustbox}

\newcommand{\RN}[1]{%
	\textup{\lowercase\expandafter{\it \romannumeral#1}}%
}
\usepackage{tabu}

\newcommand{\beq}{\vspace{0mm}\begin{equation}}
\newcommand{\eeq}{\vspace{0mm}\end{equation}}
\newcommand{\beqs}{\vspace{0mm}\begin{eqnarray}}
\newcommand{\eeqs}{\vspace{0mm}\end{eqnarray}}
\newcommand{\barr}{\begin{array}}
\newcommand{\earr}{\end{array}}

\usepackage{color, colortbl}
\definecolor{Gray}{gray}{0.93}

\usepackage{lipsum}

\usepackage{pifont}
\usepackage{makecell}

\usepackage{xcolor,amsmath}
\usepackage[linesnumbered,ruled,vlined]{algorithm2e}
\DontPrintSemicolon

\usepackage{xcolor}
\definecolor{mygreen}{HTML}{3cb44b}
            
\SetKwComment{Comment}{\color{green!50!black}\# }{}

\SetKwProg{Function}{def}{:}{}

\SetKwProg{For}{for}{:}{}
\SetKwProg{If}{if}{:}{}

\newcommand{\PredSty}[1]{\textnormal{\ttfamily\color{mygreen!90!black}#1}\unskip}

\definecolor{mygray}{gray}{.9}

\title{DarkLLM: Learning Language-Driven Adversarial Attacks with Large Language Models}

\author{
    Ye Sun$^{1*}$\quad
    Xin Wang$^{1*}$\quad
    Jiaming Zhang$^{2}$\quad
    Yifeng Gao$^{1}$\quad
    Yixu Wang$^{1}$\quad
    Yifan Ding$^{1}$ \\
\textbf{
    Qixian Zhang$^{3}$\quad
    Henghui Ding\textsuperscript{1 $\dagger$}\quad
    Xingjun Ma\textsuperscript{1 $\dagger$}\quad
    Yu-Gang Jiang\textsuperscript{1}
}
\\
$^{1}$Fudan University \quad
$^{2}$Nanyang Technological University \quad
$^{3}$Tongji University \\
\href{https://github.com/sunye23/DarkLLM}{\textbf{\texttt{https://github.com/sunye23/DarkLLM}}}
}

\hypersetup{
  colorlinks = true, 
  linkcolor = blue,  
  citecolor = green, 
  urlcolor = cyan     
}

\begin{document}

\maketitle


\begin{abstract}
\label{sec:abstract}
While vision and multimodal foundation models underpin critical tasks from perception to complex reasoning, they remain highly vulnerable to adversarial attacks. 
However, traditional adversarial attacks are typically limited to single, predefined objectives, tightly coupling each attack to a specific model or task, which restricts their scalability and flexibility in real-world scenarios. 
In this work, we present \textbf{DarkLLM}, a novel attack framework that trains an LLM to translate natural-language attack instructions into latent attack vectors, which are then decoded into visual adversarial perturbations. 
By leveraging natural-language instruction tuning, DarkLLM not only unifies targeted, untargeted, segmentation, and multi-model attacks within a single framework, but also achieves flexible and controllable adversarial generation, enabling each instruction to produce a perturbation that induces desired behaviors across heterogeneous models. 
Through extensive experiments across 4 tasks, 13 datasets, and 15 models, we demonstrate that DarkLLM with only 1B parameters can follow attacker instructions and generate highly effective attacks against CLIP, SAM, and frontier LLMs, revealing a systemic vulnerability in modern foundation models.
\end{abstract}
    
\section{Introduction}
\label{sec:introduction}
Large vision and multimodal foundation models~\cite{li2023blip,liu2023visual} have emerged as the central interface for perceiving and reasoning over visual content, underpinning a wide spectrum of applications such as image retrieval~\cite{radford2021learning}, segmentation~\cite{kirillov2023segment}, and multimodal assistants~\cite{achiam2023gpt,team2023gemini}. 
Yet their impressive capabilities are shadowed by a long-standing weakness: \emph{small, often imperceptible adversarial perturbations can reliably induce incorrect or even adversary-specified model behaviors}~\cite{szegedy2013intriguing,goodfellow2014explaining}.
Consequently, understanding and stress-testing the adversarial robustness of large-scale vision and multimodal models has become a pressing and practically relevant safety problem~\cite{madry2017towards,ma2026safety}.

\begin{figure*}[t]
  \centering
\includegraphics[width=0.99\linewidth]{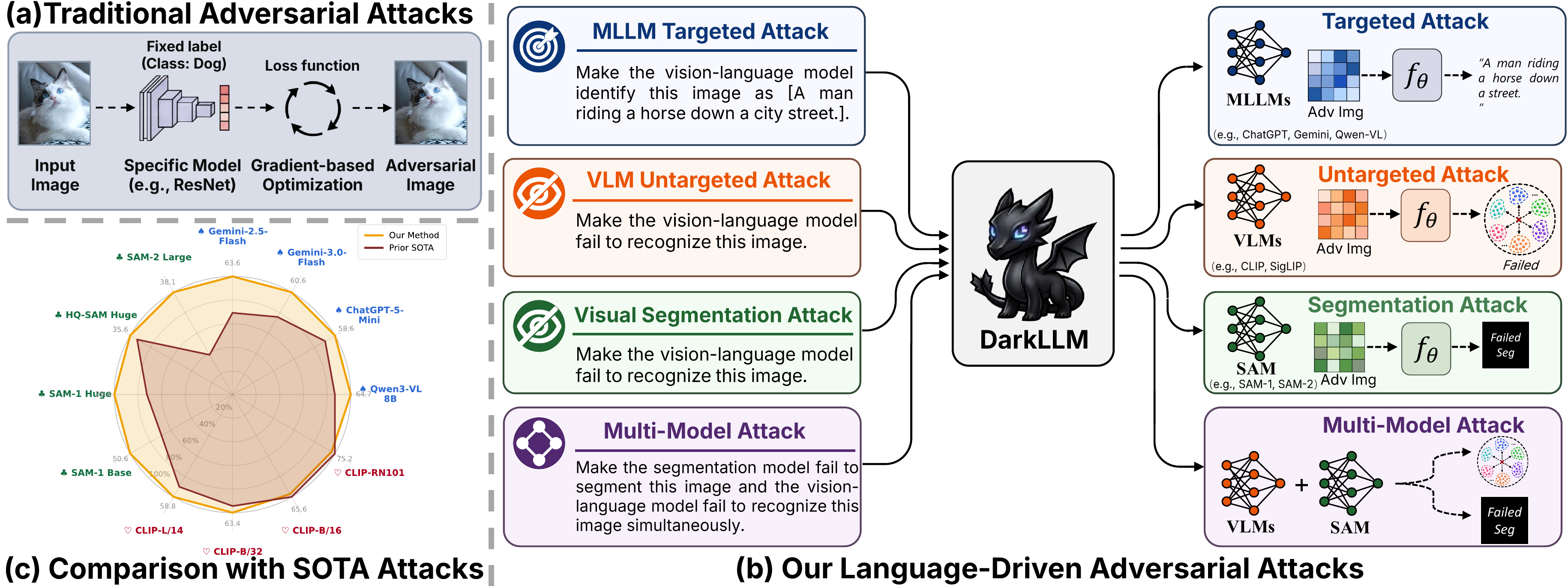}
\vspace{-1.5mm}
\caption{(a) An illustration of traditional adversarial attacks.
(b) DarkLLM translates natural-language instructions into customized visual perturbations, enabling unified and controllable adversarial attacks.
(c) DarkLLM achieves strong attack success rates across diverse models.}
\label{fig:teaser}
   \vspace{-7.5mm}
\end{figure*}

Over the past decade, adversarial research has been dominated by gradient-based optimization~\cite{goodfellow2014explaining,madry2017towards} and generative adversarial modeling~\cite{goodfellow2014generative,xiao2018generating}, revealing pervasive vulnerabilities across both classical architectures and modern foundation models~\cite{moosavi2017universal,zhao2023evaluating}. 
Yet two fundamental challenges persist: \emph{transferability} and \emph{scalability}. 
First, an effective attack should be transferable across heterogeneous models and tasks—often even jointly, as contemporary systems routinely integrate recognition, segmentation, and multimodal reasoning modules~\cite{zhang2022towards,zhou2024darksam,zhao2023evaluating}.
Equally importantly, a practical attack framework should be scalable, allowing different adversarial intents to be flexibly specified without requiring objective-specific pipeline redesign or repeated optimization~\cite{zhang2025anyattack,fang2024clip}.
However, existing approaches struggle to satisfy these requirements simultaneously. 
Most untargeted attacks that aim to induce arbitrary errors are tailored to specific architectures or foundation models (e.g., ViT~\cite{wei2022towards,mahmood2021robustness,shao2021adversarial}, CLIP~\cite{zhang2022towards,zhou2023advclip,fang2025one}, or SAM~\cite{zhou2024darksam,zhang2023attack,lu2024unsegment}), while representative targeted attacks, such as SU~\cite{wei2023enhancing}, SASD~\cite{wu2024improving}, TTAA~\cite{wang2023towards}, TTP~\cite{naseer2021generating}, SSA-CWA~\cite{dong2023robust}, and M-Attack~\cite{li2025frustratingly}, steer model outputs toward adversary-specified targets but typically follow a rigid ``one-target, one-optimization'' paradigm, requiring re-optimization for each new adversarial objective.
These limitations raise a critical question: 
\emph{Is it possible to design a new adversarial attack framework that is transferable and scalable across models and objectives?}

In parallel, the emergence of Large Language Models (LLMs)~\cite{brown2020language} has fundamentally reshaped how complex tasks are specified and solved. LLMs excel at interpreting semantically rich, compositional instructions and translating them into structured, domain-specific outputs, enabling breakthroughs across visual understanding~\cite{liu2023visual,xie2024show,sun2026sama}, generation~\cite{tian2024visual,chen2025janus}, and medical analysis~\cite{singhal2025toward,peng2023study}. Within the safety domain, LLMs have already been exploited to generate jailbreak prompts~\cite{liu2023autodan,liu2024autodan}, automate safety evaluations~\cite{ying2026safebench}, and design textual backdoors~\cite{li2025autobackdoor,feng2026backdooragent}.
Yet, existing work largely remains confined to the textual domain, where LLMs are used to generate adversarial or malicious text, rather than pixel-level adversarial perturbations, motivating an intriguing problem:
\emph{Can the semantic reasoning capabilities of LLMs be leveraged for visual adversarial generation?}

In this work, we propose a novel language-driven attack framework, \textbf{DarkLLM}, which leverages natural-language instructions as the interface for adversarial generation. Figure~\ref{fig:teaser} highlights the distinction between DarkLLM and existing attack strategies. Unlike conventional attacks, DarkLLM reformulates adversarial generation as a language-conditioned synthesis problem, enabling each instruction to be translated into a visual perturbation that induces desired adversarial behaviors across heterogeneous foundation models. 
Specifically, our approach trains an LLM to act as an adversarial controller, mapping semantically rich attack instructions into latent attack representations, which are then decoded by a lightweight perturbation generator into pixel-level adversarial noise. Through large-scale instruction tuning, DarkLLM learns to associate different attack intents with corresponding generators, allowing targeted, untargeted, segmentation, and multi-model attacks to be flexibly generated within a unified framework. We systematically analyze several key factors for improving attack effectiveness, including prompt ensemble strategies, loss function design, surrogate model selection, and training strategies. Finally, we conduct extensive experiments across 4 tasks and 15 models (CLIP~\cite{radford2021learning}, SAM~\cite{kirillov2023segment,ravi2024sam}, Gemini~\cite{team2023gemini}, ChatGPT~\cite{achiam2023gpt}), demonstrating that DarkLLM achieves strong adversarial effectiveness across all evaluated foundation models. 

Our main contributions are summarized as follows:
\begin{itemize}
    \item We propose \textbf{DarkLLM}, a novel language-driven adversarial attack  framework that reformulates visual adversarial generation as an instruction-conditioned synthesis problem, enabling diverse attack objectives to be flexibly specified through natural language. 

    \item We develop a unified adversarial generation paradigm that leverages LLM-based adversarial controllers and modular perturbation generators to jointly support targeted, untargeted, segmentation, and multi-model attacks within a single framework.

    \item We conduct extensive experiments across 4 tasks and 15 models, covering CLIP, SAM, and frontier multimodal LLMs. Both qualitative and quantitative results demonstrate that DarkLLM achieves strong adversarial effectiveness and robustness against common image corruption defenses.
\end{itemize}

\section{Related Work}
\label{sec:related_work}
\noindent\textbf{Conventional Adversarial Attacks.}\quad
The vulnerability of deep neural networks to adversarial examples remains a foundational challenge in machine learning~\cite{szegedy2013intriguing}. Traditional research divides into white-box attacks, which leverage full model access (e.g., gradients) to craft perturbations via optimization methods like FGSM~\cite{goodfellow2014explaining} and PGD~\cite{madry2017towards}, and black-box attacks, which operate under restricted access. Black-box strategies typically rely on query-based estimation~\cite{ilyas2018black, andriushchenko2020square} or transfer-based generation using surrogate models~\cite{dong2018boosting, xie2019improving,huang2025x, li2025frustratingly}. While effective on specific tasks, conventional methods typically operate under a fixed optimization framework tailored to a single target. This constraint severely limits their scalability across the diverse and dynamic task spectrum of modern foundation models.

\noindent\textbf{Adversarial Attacks on Foundation Models.}\quad
The rise of large vision foundation models has introduced architectures with unprecedented generalization~\cite{radford2021learning, kirillov2023segment, liu2023visual}. Yet, recent studies confirm that this capability does not equate to robustness; these models remain highly susceptible to adversarial perturbations~\cite{zhang2022towards,zhao2023evaluating}.
Existing research has systematically exposed these vulnerabilities across different domains: adversarial attacks targeting CLIP's modality alignment include specific methods like Co-Attack~\cite{zhang2022towards}, AdvCLIP~\cite{zhou2023advclip}, C-PGC~\cite{fang2025one} and X-Transfer~\cite{huang2025x}; vulnerabilities in SAM's segmentation boundaries have been exploited by attacks such as Attack-SAM~\cite{zhang2023attack}, DarkSAM~\cite{zhou2024darksam}, UAD~\cite{lu2024unsegment}, and UAP-SAM2~\cite{zhou2025vanish}; and the multimodal reasoning of large VLMs has been compromised by approaches like AttackVLM~\cite{zhao2023evaluating}, JailbreakV-28K~\cite{luo2024jailbreakv}, AnyAttack~\cite{zhang2025anyattack}, FOA-Attack~\cite{jia2025adversarial}, and M-Attack~\cite{li2025frustratingly}.
However, these efforts have largely operated in isolation, treating each model and task as a distinct optimization problem with specialized objectives. In contrast, our work aims to unify diverse attack settings within a single language-driven framework, enabling transferable adversarial generation across heterogeneous architectures.

\section{Proposed Method}
\label{sec:method}
\paragraph{Threat Model.}
We focus on a transfer-based black-box attack scenario. The adversary aims to synthesize an adversarial example $x_{\text{adv}}$ using accessible surrogate models $\mathcal{F}_s$, such that it successfully deceives unseen target models $\mathcal{F}_t$.
We formulate this process as an \textit{Language-guided adversarial attack}. Unlike traditional methods bound to fixed labels, our framework flexibly constructs diverse adversarial objectives via natural language. Let $\mathcal{D} = \{(x, y)\}$ be a dataset of clean images and annotations. The goal is to craft $x_{\text{adv}} = x + \delta$ that misleads $\mathcal{F}_t$, where the specific adversarial goal is dynamically instantiated from a user instruction $I$. For instance, given an instruction targeting a multimodal large language model, the objective is to align $x_{\text{adv}}$ with a malicious target caption $y_t$~\cite{zhang2025anyattack}; conversely, for a promptable segmentation model like SAM, the instruction directs the attack to suppress all mask predictions (i.e., ``segment nothing''), regardless of visual prompts~\cite{zhou2024darksam}. 

\begin{figure*}[t]
\begin{center}
\includegraphics[width=1.0\linewidth]{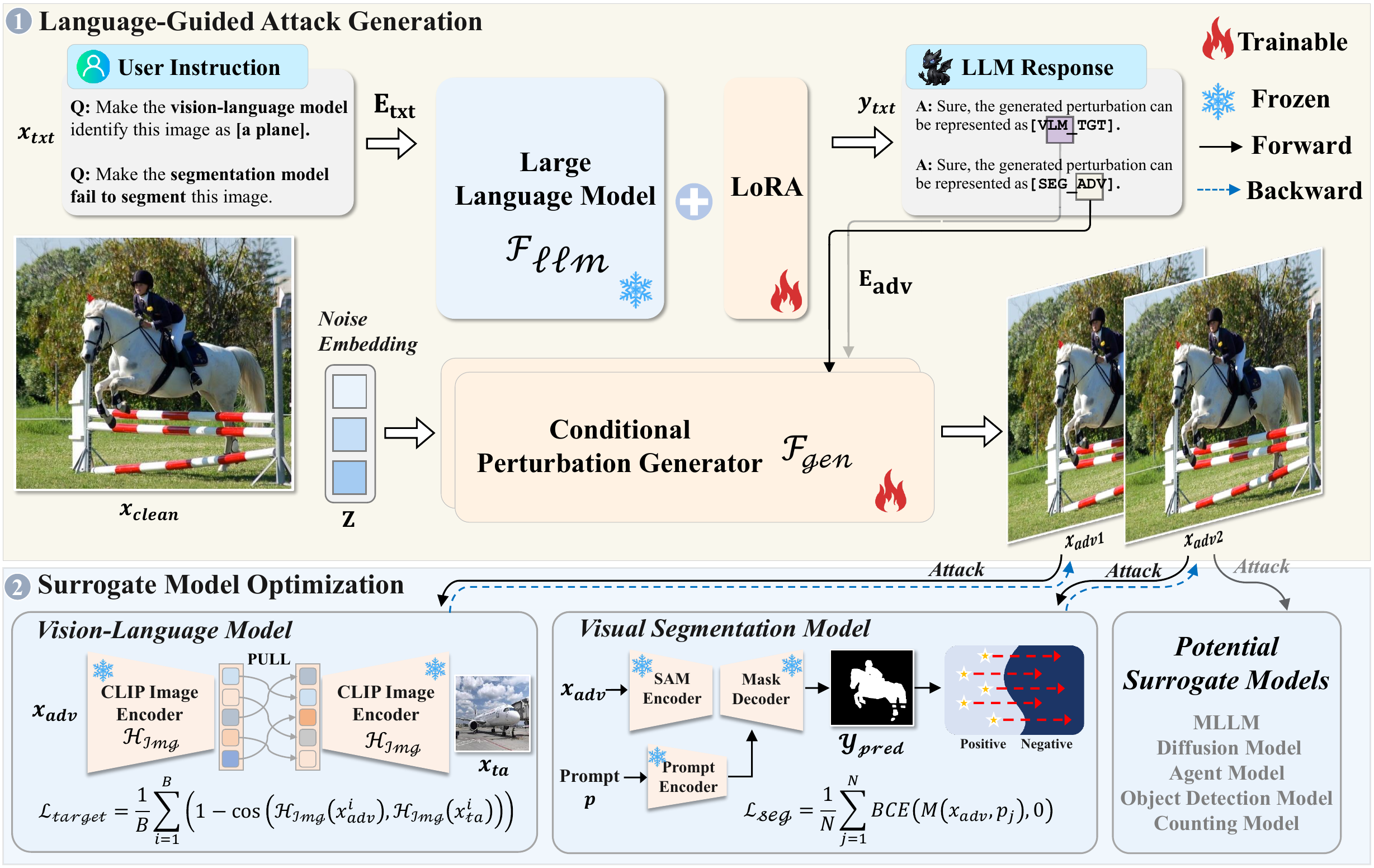}
\end{center}
\vspace{-1em}
\caption{
    The framework of DarkLLM consists of two main stages.
    \textbf{1)} Language-Guided Attack Generation, where a LLM controller interprets a user's instruction and directs a conditional generator to synthesize the perturbation. 
\textbf{2)} Surrogate Model Optimization, where the generated adversarial example is optimized using a diverse suite of surrogate models to ensure transferability.
}
\label{fig:framework}
\vspace{-1em}
\end{figure*}
\subsection{DarkLLM Framework}
\label{sec:overview}
\paragraph{Embedding as Attack.}
Conventional adversarial attacks typically formulate adversarial objectives as task-specific optimization targets, tightly coupling each attack behavior with a dedicated optimization pipeline. While effective for individual settings, such formulations make attacks difficult to adapt across heterogeneous models, tasks, and objectives, resulting in fragmented methods separately designed for classification, segmentation, or targeted generation. In contrast, DarkLLM reformulates adversarial generation as a language-conditioned synthesis process, where natural-language instructions are interpreted by an LLM and mapped into high-level latent attack embeddings that condition downstream perturbation generators, enabling diverse attack behaviors to be flexibly generated within a unified framework.

To this end, we propose the \textit{Embedding as Attack} paradigm, a novel approach that repurposes a Large Language Model (LLM) to act as a versatile and language-guided adversarial controller. The overall pipeline is illustrated in Figure~\ref{fig:framework}. Specifically, central to this paradigm is the introduction of a new set of special control tokens (e.g., \texttt{[VLM\_TGT]}, \texttt{[SEG\_ADV]}) into the LLM’s vocabulary, where each token serves as a dedicated trigger for invoking a corresponding adversarial generator to synthesize a specific type of perturbation.
Our framework begins when the core LLM controller, $\mathcal{F}_\text{LLM}$, receives the language embeddings $\mathbf{E}_\text{txt}$ of a user's instruction $x_{txt}$. Based on its interpretation of the instruction, the controller decides whether to generate an adversarial attack. If so, it strategically embeds the specific control token within its textual response $y_{txt}$:
\begin{equation}
    y_{txt} = \mathcal{F}_\text{LLM}(\mathbf{E}_\text{txt}).
\end{equation}

The presence of the control token in the generated response serves as a trigger, activating the attack generation module. To be specific, we extract the final-layer hidden state $\mathbf{\tilde{h}}_{\text{adv}}$ corresponding to the control token, which represents the LLM's latent understanding of the required attack. This latent attack embedding is then projected through a two-layer MLP, $\gamma$, to form a refined attack signal, $\mathbf{E}_{\text{adv}}$. This signal conditions a trainable perturbation generator, $\mathcal{F}_{\text{gen}}$, guiding it to synthesize the appropriate adversarial perturbation $\delta_{\text{adv}}$. The generation process is formulated as:
\begin{equation}
\begin{aligned}
    \mathbf{E}_{\text{adv}} &= \gamma(\mathbf{\tilde{h}}_{\text{adv}}), \delta_{\text{adv}}=\mathcal{F}_{\text{gen}}(\mathbf{E}_{\text{adv}}).
\end{aligned}
\end{equation}
Finally, the perturbation is added to a clean image $x_{\text{clean}}$ to produce the final adversarial example, $x_{\text{adv}} = x_{\text{clean}} + \delta_{\text{adv}}$.
\\\textbf{Conditional Perturbation Generator.}\quad 
\label{sec:generator_arch}
\begin{wrapfigure}[18]{R}{0.25\textwidth}
\centering
\vspace{-6pt}
\includegraphics[width=0.25\textwidth]{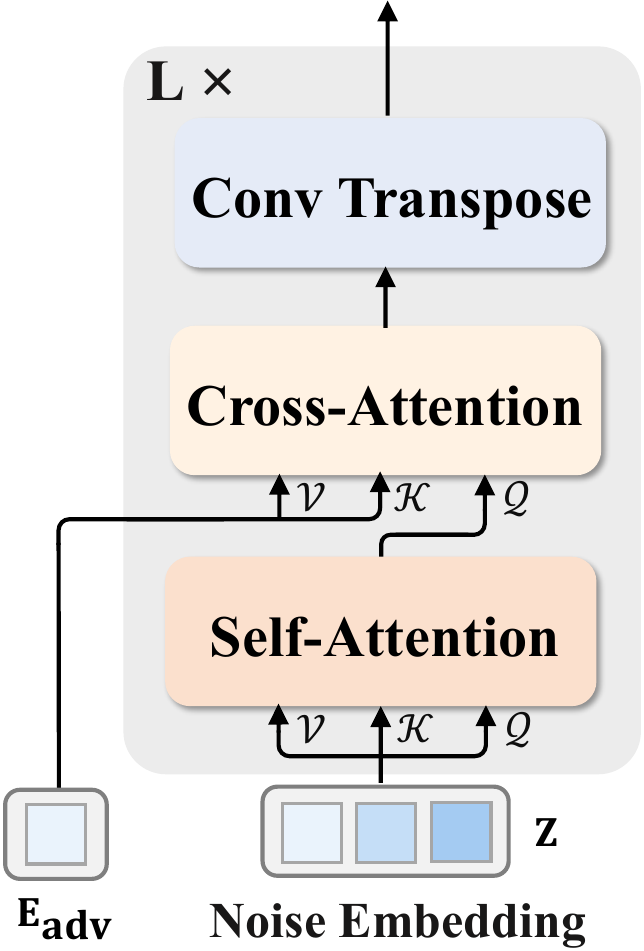}
\caption{A brief illustration of our conditional noise generator. }
\label{fig:noise_generator}
\end{wrapfigure}
To translate the high-level adversarial representation $\mathbf{E}_{\text{adv}}$ into an image-space perturbation $\delta_{\text{adv}}$, we employ a decoder network that progressively refines a set of latent noise embeddings through a series of attention blocks, as illustrated in Figure~\ref{fig:noise_generator}.
Specifically, the generator $\mathcal{F}_{\text{gen}}$, parameterized by $\theta_\text{G}$, accepts two inputs: the latent attack embedding $\mathbf{E}_{\text{adv}} \in \mathbb{R}^{1 \times C}$ derived from the LLM $\mathcal{F}_\text{LLM}$ and a random noise embedding $\mathbf{Z} \in \mathbb{R}^{3 \times 3 \times C}$ sampled from a standard normal distribution, $\mathbf{Z} \sim \mathcal{N}(0, I)$, where C denotes the number of input channels. The adversarial perturbation is generated by $\mathcal{F}_{\text{gen}}$ as follows: 
\begin{gather}
\delta_\text{adv} = \mathcal{F}_{\text{gen}}(\mathbf{E}_{\text{adv}}, \mathbf{Z}; \theta_G) \quad \text{s.t.} \quad \|\delta_\text{adv}\|_{\infty} \leq \epsilon.
\label{eq:eq4}
\end{gather}

To infuse the noise embeddings with the specific adversarial intent derived from the LLM, we employ the cross-attention mechanism as the primary vehicle for information injection. At each building block of our generator, the noise embeddings $\mathbf{Z}$ are used as the query, whereas the latent attack embedding $\mathbf{E}_{\text{adv}}$ from the LLM serves as both the key and the value. This allows the latent attack embedding to guide the refinement of the noise embeddings, effectively steering the synthesis process towards the user's specified adversarial goal. The operation can be formulated as:
\begin{equation}
\label{eq:cross_attention}
\mathrm{Softmax}((\mathbf{Z}\mathbf{W}_Q)\times(\mathbf{E}_{\text{adv}}\mathbf{W}_K)^\top/\sqrt{C})\times\mathbf{E}_{\text{adv}}\mathbf{W}_V
\end{equation}
\noindent where $\mathbf{W}_Q, \mathbf{W}_K, \mathbf{W}_V$ denote the linear projection layer. In addition to cross-attention, each block of generator incorporates a self-attention layer to capture global dependencies, followed by transposed convolutions for spatial upsampling. Finally, we use bilinear interpolation to resize the generated perturbation to the target image spatial dimensions and apply a Tanh activation function to ensure the final perturbation $\delta_{\text{adv}}$ is bounded by $\|\delta_{\text{adv}}\|_\infty \le \epsilon$.

\subsection{Training the DarkLLM}
\label{sec:training}
We employ two representative classes of large vision foundation models as surrogates for optimizing DarkLLM: the coarse-grained vision-language model, CLIP~\cite{radford2021learning}, and the fine-grained promptable visual segmentation model, SAM~\cite{kirillov2023segment}. 
\\\textbf{Optimization for Fine-Grained Segmentation.}\quad
To extend our framework to fine-grained visual segmentation, we incorporate SAM as a key surrogate model. 
However, directly attacking SAM poses two challenges: 
(1) \textit{Prompt Transferability:} perturbations optimized for a specific object or prompt type (e.g., a point prompt) often fail to generalize to other prompts; 
(2) \textit{Backbone Transferability:} perturbations optimized on one SAM backbone exhibit degraded transfer to other backbones. To improve prompt transferability, we adopt a multi-prompt ensemble optimization strategy. For each image, we randomly sample a diverse set of prompts $\mathcal{P}=\{p_1,\ldots,p_N\}$ covering both point and box prompts, and optimize a single perturbation to suppress the predicted masks under all sampled prompts, thereby encouraging prompt-agnostic ``segment nothing'' behavior. 
Nevertheless, backbone transfer remains non-trivial. Through careful ablation studies, we find that poor cross-backbone transfer largely stems from widely used regression-style losses (e.g., Mean Squared Error), which over-emphasize matching backbone-specific activation magnitudes and cause perturbations to overfit a particular SAM variant. 
To address this issue, we instead optimize a binary cross-entropy (BCE) objective that pushes mask predictions toward the all-background decision boundary, yielding gradients that are more semantically aligned and less sensitive to backbone-specific activations. Concretely, we define:
\begin{equation}
\mathcal{L}_{\text{seg}} = \frac{1}{N} \sum_{j=1}^{N} 
\mathrm{BCE}\!\left( M(x_{\text{adv}}, p_j), \mathbf{0} \right),
\label{eq:sam_bce}
\end{equation}
where $M(x_{\text{adv}}, p_j)$ denotes the mask logits predicted by SAM for the adversarial image $x_{\text{adv}}$ under prompt $p_j$. We also include a lightweight MobileSAM~\cite{zhang2023faster} as an additional surrogate to improve transferability.
\\\textbf{Optimization for Vision-Language Models.}\quad
Achieving high cross-model transferability is paramount when attacking the diverse family of CLIP models. To this end, we adopt a surrogate scaling strategy inspired by X-Transfer~\cite{huang2025x}. Specifically, during distributed training, each computational rank (e.g., a GPU) is assigned a different surrogate from a large pool of CLIP models with varying architectures and pre-training datasets, ensuring that gradients from a diverse set of models are jointly incorporated and thereby enhancing overall transferability. Our framework supports both untargeted and targeted attacks, with the specific attack objective dynamically determined by the user’s instruction. In both cases, optimization is performed in the CLIP embedding space using cosine similarity. For untargeted attacks, the objective is to disrupt the original image–text alignment by minimizing the similarity between the adversarial image embedding and the benign image embedding. For targeted attacks, the objective is reversed: to maximize the similarity between the adversarial image embedding and the embedding of a target image $x_{\text{ta}}$. These objectives are formalized by the following loss functions:
\begin{align}
\mathcal{L}_{\text{untarget}} &= \frac{1}{B} \sum_{i=1}^{B} 
\cos\!\left(\mathcal{H}_{\text{img}}(x^{i}_{\text{adv}}), 
\mathcal{H}_{\text{img}}(x^{i}_{\text{clean}})\right),
\label{eq:clip_loss_untargeted} \\[4pt]
\mathcal{L}_{\text{target}} &= \frac{1}{B} \sum_{i=1}^{B} 
\left( 1 - \cos\!\left(\mathcal{H}_{\text{img}}(x^{i}_{\text{adv}}), 
\mathcal{H}_{\text{img}}(x^{i}_{\text{ta}})\right) \right),
\label{eq:clip_loss_targeted}
\end{align}
\noindent where $\mathcal{H}_{\text{img}}$ is the image encoder, and $B$ is the batch size. 

\paragraph{Decoupled Progressive Training.}
Unifying heterogeneous attack types within a single instruction-following framework is non-trivial, as different attacks require different levels of semantic understanding and exhibit highly imbalanced convergence behaviors. 
In our early joint-training experiments, we observed that untargeted attacks converge much faster than language-sensitive targeted attacks, and their rapidly updated parameters dominate optimization, significantly degrading targeted performance. 
To address this issue, we adopt an effective \emph{decoupled progressive} training strategy. We assign unique control tokens (e.g., \texttt{[VLM\_TGT]}, \texttt{[VLM\_ADV]}, \texttt{[SEG\_ADV]}, \texttt{[MULTI\_ADV]}) and dedicated noise generators to each attack category, preventing gradient interference across incompatible objectives. Training proceeds in two stages. \emph{(1) Targeted Instruction Tuning:} We first instruction-tune the LLM controller via LoRA on targeted QA-style attack data until convergence, enabling precise parsing and execution of semantically rich attack commands. \emph{(2) Unified Attack Training:} We then initialize the controller with stage-1 weights and train all attack types jointly, while freezing the targeted generator to preserve its capability and learning additional untargeted and multi-model attacks through their respective generators. 
This decoupled progression yields stable and efficient training, while offering modularity and scalability for incorporating new attack types.
\section{Experiments}
\label{sec:experiments}
\subsection{Experimental Setup}
\textbf{Implementation Details.}\quad 
We implement DarkLLM based on the XTuner~\cite{2023xtuner} codebase. The LLM is initialized from InternVL2.5-1B~\cite{chen2024expanding} and fine-tuned using LoRA~\cite{hu2022lora} on 8 NVIDIA H200 GPUs. Training follows our decoupled progressive strategy. 
In stage~1, we jointly train on ImageNet~\cite{deng2009imagenet} for 55 epochs and CC3M~\cite{sharma2018conceptual} for 10 epochs using the AdamW optimizer with an initial learning rate of $1\times10^{-4}$. The per-GPU batch size is set to 128. In stage~2, we further fine-tune the model on ImageNet for 8 epochs with a batch size of 64. The overall training objective is defined as a weighted sum of individual surrogate losses:
\begin{equation}
\mathcal{L}_{\text{total}}
= \lambda_{\text{t}}\mathcal{L}_{\text{txt}}
+ \lambda_{\text{s}}\mathcal{L}_{\text{seg}}
+ \lambda_{\text{ta}}\mathcal{L}_{\text{target}}
+ \lambda_{\text{un}}\mathcal{L}_{\text{untarget}},
\label{eq:sam_loss}
\end{equation}
The hyperparameters $\lambda_{\text{t}}, \lambda_{\text{s}}, \lambda_{\text{ta}},$ and $\lambda_{\text{un}}$ are set to 1.0, 10.0, 10.0, and 8.0, respectively, to balance different attack objectives. We apply an $L_{\infty}$-norm constraint on all perturbations. Specifically, we set $\epsilon = 12/255$ for untargeted VLM attacks following~\cite{huang2025x,fang2025one}, $\epsilon = 16/255$ for targeted MLLM attacks following~\cite{zhang2025anyattack,li2025frustratingly,jia2025adversarial}, and $\epsilon = 10/255$ for SAM attacks following~\cite{zhou2024darksam,zhou2025vanish}.

\noindent \textbf{Baselines, Tasks, and Datasets.}\quad Please refer to the Appendix \ref{sec:supp-exps}. for details due to page limitations.

\newcommand{\colorrowmulti}[3]{%
  \rowcolor{#1}%
  \multirow{2}{*}{#2} & #3 \\[-1em]
  \rowcolor{#1}%
  & #3 \\%
}
\newcommand*{\belowrulesepcolor}[1]{%
  \noalign{%
    \kern-\belowrulesep 
    \begingroup 
      \color{#1}%
      \hrule height\belowrulesep 
    \endgroup 
    \vspace{-0.03mm}
  }%
} 
\newcommand*{\aboverulesepcolor}[1]{%
  \noalign{%
  \vspace{-0.03mm}
    \begingroup 
      \color{#1}%
      \hrule height\aboverulesep 
    \endgroup 
    \kern-\aboverulesep 
  }%
}
\begin{table*}[t]
\centering
\caption{Non-targeted ASR (\%) results for zero-shot classification, image-text (I-T) retrieval, and image captioning tasks. The best results are \textbf{boldfaced}, and the second-best are \underline{underlined}. \textit{Results for classification and retrieval are averaged over 4 black-box CLIP encoders.}
}
\begin{adjustbox}{width=0.99\linewidth}
\begin{tabular}{@{}l|c|cccccccc|cc@{}|cc}
\toprule[1.5pt]
\multirow{2}{*}{\textbf{Method}} & \multirow{2}{*}{\textbf{Variant}} & \multicolumn{8}{c|}{\textbf{Zero-Shot Classification}} & \multicolumn{2}{c|}{\textbf{I-T Retrieval}} & \multicolumn{2}{c}{\textbf{Image Captioning}}\\ \cmidrule(l){3-14}
 &  & \texttt{\textbf{C-10}} & \texttt{\textbf{C-100}} & \texttt{\textbf{Food}} & \texttt{\textbf{GTSRB}} & \texttt{\textbf{ImageNet}} & \texttt{\textbf{Cars}} & \texttt{\textbf{STL}} & \texttt{\textbf{Avg}} & \texttt{\textbf{TR@1}} & \texttt{\textbf{IR@1}} & \texttt{\textbf{MSCOCO}} & \texttt{\textbf{Flickr30k}} \\ \midrule[0.75pt]

\multirow{2}{*}{GD-UAP~\cite{mopuri2018generalizable}} & Seg & 63.2 & 78.2 & 36.1 & 58.2 & 22.2 & 19.9 & 9.8 & 41.0 & 26.3 & 21.8  & 8.1 & 6.5 \\ 
 & CLS & 66.5 & 80.4 & 55.2 & 61.7 & 29.8 & 28.8 & 20.1 & 48.9 & 29.5 & 24.7  & 10.3 & 8.6 \\ 
 \midrule[0.75pt]

\multirow{2}{*}{TRM-UAP~\cite{liu2023trm}} & GoogleNet & 73.3 & 84.0 & 60.7 & 71.1 & 33.0 & 30.5 & 24.7 & 53.9 & 30.0 & 25.7 & 6.6 & 10.0 \\ 
 & RN152 & 58.0 & 73.9 & 54.4 & 63.7 & 27.9 & 27.8 & 19.1 & 46.4 & 26.1 & 22.1 & 2.8 & 7.3 \\ 
 \midrule[0.75pt]

\multirow{2}{*}{Meta-UAP~\cite{weng2024learning}} & Ensemble & 82.7 & \underline{93.2} & 59.8 & 64.9 & 39.4 & 39.2 & 27.7 & 58.1 & 37.1 & 32.1 & 16.3 & 15.0 \\ 
 & Ensemble-Meta & 72.3 & 88.5 & 53.3 & 59.3 & 35.8 & 34.6 & 22.1 & 52.2 & 35.3 & 29.3 & 11.3 & 12.7 \\ 
 \midrule[0.75pt]

\multirow{2}{*}{ETU~\cite{zhang2024universal}} & RN50-Flickr & 44.2 & 64.0 & 32.7 & 46.9 & 21.8 & 21.6 & 14.5 & 35.1 & 18.0 & 15.5 & 7.4 & 5.2 \\
 & ViT-B/16-Flickr & 76.2 & 89.7 & 67.8 & 68.5 & 51.5 & 53.4 & 40.7 & 63.9 & 49.0 & 46.5 & 17.8 & 16.2 \\ \midrule[0.75pt]

\multirow{2}{*}{C-PGC~\cite{fang2025one}}  
& RN101-COCO & 34.8 & 60.5 & 37.8 & 47.9 & 31.8 & 30.3 & 21.0 & 37.7 & 33.2 & 26.3 & 1.46 & 2.1 \\
& ViT-B/16-COCO & 74.4 & 90.3 & 65.1 & 72.2 & 53.0 & 56.7 & 36.8 & \underline{64.0} & 51.6 & 52.7 & 11.5 & 10.5 \\ \midrule[0.75pt]
 
\multirow{2}{*}{X-Transfer~\cite{huang2025x}} & Vanilla & 69.1 & 85.8 & 60.6 & 64.7 & 37.2 & 34.3 & 19.4 & 53.0 & 33.7 & 30.4 & 9.6 & 11.4 \\ 
 & Base & \underline{86.2} & \textbf{97.3} & \textbf{83.6} & \textbf{86.2} & \underline{63.4} & \underline{69.7} & \textbf{49.3} & \textbf{76.5} & \underline{59.7} & \underline{55.2} & \textbf{21.0} & \underline{18.5} \\
\midrule[0.75pt]
\belowrulesepcolor{blue!10!}
\rowcolor{blue!10} 
DarkLLM (Ours) & Ensemble & \textbf{87.4} & \textbf{97.3} & \underline{83.5} & \underline{85.8} & \textbf{64.7} & \textbf{71.7} & \underline{45.0} & \textbf{76.5} & \textbf{62.9} & \textbf{59.3} & \underline{19.3} & \textbf{19.0} \\ 
\aboverulesepcolor{blue!10!}
\bottomrule[1.5pt]
\end{tabular}
\end{adjustbox}
\vspace{-4mm}
\label{tab:untargeted_vlm}
\end{table*}

\begin{table*}[t]
\centering
\vspace{-2mm}
\caption{Targeted ASR (\%) and AvgSim results for image captioning tasks across different commercial MLLMs. The best results among universal methods are \textbf{boldfaced}, and the second-best results are \underline{underlined}.}
\resizebox{0.99\linewidth}{!}{ 
\begin{tabular}{l|c|cc|cc|cc|cc|cc|cc}
\toprule[1.5pt]
& 
& \multicolumn{2}{c|}{\textbf{Qwen3-VL-8B}} 
& \multicolumn{2}{c|}{\textbf{GPT-4o}} 
& \multicolumn{2}{c|}{\textbf{GPT-4.1}} 
& \multicolumn{2}{c|}{\textbf{GPT-5-mini}} 
& \multicolumn{2}{c|}{\textbf{Gemini-2.5-Flash}}
& \multicolumn{2}{c}{\textbf{Gemini-3-Flash}} \\
\cmidrule{3-4} \cmidrule{5-6} \cmidrule{7-8} \cmidrule{9-10} \cmidrule{11-12} \cmidrule{13-14} 
\multirow{-2}{*}{\textbf{Method}} & \multirow{-2}{*}{\textbf{Variant}} & \texttt{\textbf{ASR}} & \texttt{\textbf{AvgSim}} & \texttt{\textbf{ASR}} & \texttt{\textbf{AvgSim}} & \texttt{\textbf{ASR}} & \texttt{\textbf{AvgSim}} & \texttt{\textbf{ASR}} & \texttt{\textbf{AvgSim}} & \texttt{\textbf{ASR}} & \texttt{\textbf{AvgSim}} & \texttt{\textbf{ASR}} & \texttt{\textbf{AvgSim}} \\
    \hline
    \rowcolor{black!5} \multicolumn{14}{c}{\textit{Sample-wise Adversarial Methods}} \\
    \hline
\multirow{3}{*}{AttackVLM~\cite{zhao2023evaluating}}
& B/16 & 3.0 & 0.02 & 3.0 & 0.01 & 3.0 & 0.02 & 0.0 & 0.00 & 1.0 & 0.01 & 0.0 & 0.00 \\
& L/14  & 4.0 & 0.03 & 3.0 & 0.02 & 1.0 & 0.01 & 1.0 & 0.00 & 2.0 & 0.02 & 1.0 & 0.00 \\
& Ensemble & 3.0 & 0.02 & 2.0 & 0.02 & 3.0 & 0.01 & 3.0 & 0.01 & 2.0 & 0.02 & 2.0 & 0.01\\
    \hline
\multirow{3}{*}{M-Attack~\cite{li2025frustratingly}}
& B/16 & 31.0 & 0.15 & 43.0 & 0.19 & 50.0 & 0.20 & 26.0 & 0.12 & 25.0 & 0.11 & 21.0 & 0.08 \\
& L/14  & 52.0 & 0.28 & 65.0 & 0.31 & 66.0 & 0.31 & 34.0 & 0.16 & 53.0 & 0.25 & 42.0 & 0.17 \\
& Ensemble & 87.0 & 0.47 & 86.0 & 0.47 & 89.0 & 0.51 & 63.0 & 0.29 & 77.0 & 0.38 & 69.0 & 0.32\\
    \hline
    \rowcolor{black!5} \multicolumn{14}{c}{\textit{Universal Adversarial Methods}} \\
    \hline
\multirow{4}{*}{AnyAttack~\cite{zhang2025anyattack}}
& coco-bi & \underline{56.0} & \underline{0.24} & 60.0 & 0.22 & 59.0 & \underline{0.25} & 53.0 & \underline{0.22} & 44.0 & 0.19 & 46.0 & 0.18 \\
& coco-cos & 5.0 & 0.02 & 7.0 & 0.04 & 4.0 & 0.03 & 5.0 & 0.04 & 7.0 & 0.04 & 8.0 & 0.03 \\
& flickr-bi & 36.0 & 0.17 & 40.0 & 0.16 & 47.0 & 0.19 & 32.0 & 0.16 & 32.0 & 0.14 & 23.0 & 0.09\\
& flickr-cos & 25.0 & 0.11 & 29.0 & 0.14 & 34.0 & 0.14 & 27.0 & 0.12 & 30.0 & 0.14 & 20.0 & 0.09\\
    \hline
\cellcolor{blue!10} & 
\cellcolor{blue!10} InternVL-1B & 
\cellcolor{blue!10} 49.5 & 
\cellcolor{blue!10} \underline{0.24} & 
\cellcolor{blue!10} \underline{64.6} & 
\cellcolor{blue!10} \underline{0.26} & 
\cellcolor{blue!10} \textbf{67.6} & 
\cellcolor{blue!10} \textbf{0.27} & 
\cellcolor{blue!10} \underline{54.5} & 
\cellcolor{blue!10} \textbf{0.24} & 
\cellcolor{blue!10} \underline{57.5} & 
\cellcolor{blue!10} \underline{0.24} & 
\cellcolor{blue!10} \underline{52.5} & 
\cellcolor{blue!10} \underline{0.22} \\

\cellcolor{blue!10} \multirow{-2}{*}{DarkLLM (Ours)} & 
\cellcolor{blue!10} InternVL-2B & 
\cellcolor{blue!10} \textbf{64.7} & 
\cellcolor{blue!10} \textbf{0.29} & 
\cellcolor{blue!10} \textbf{67.7} & 
\cellcolor{blue!10} \textbf{0.28} & 
\cellcolor{blue!10} \underline{64.6} & 
\cellcolor{blue!10} \textbf{0.27} & 
\cellcolor{blue!10} \textbf{58.6} & 
\cellcolor{blue!10} \textbf{0.24} & 
\cellcolor{blue!10} \textbf{63.6} & 
\cellcolor{blue!10} \textbf{0.25} & 
\cellcolor{blue!10} \textbf{60.6} & 
\cellcolor{blue!10} \textbf{0.23} \\
\aboverulesepcolor{blue!10!}
\bottomrule[1.5pt]
\end{tabular}
} 
\vspace{-3mm}
\label{tab:attack_mllm}
\end{table*}

\subsection{Main Results}
\begin{figure*}[!t]
\begin{center}
\includegraphics[width=0.98\linewidth]{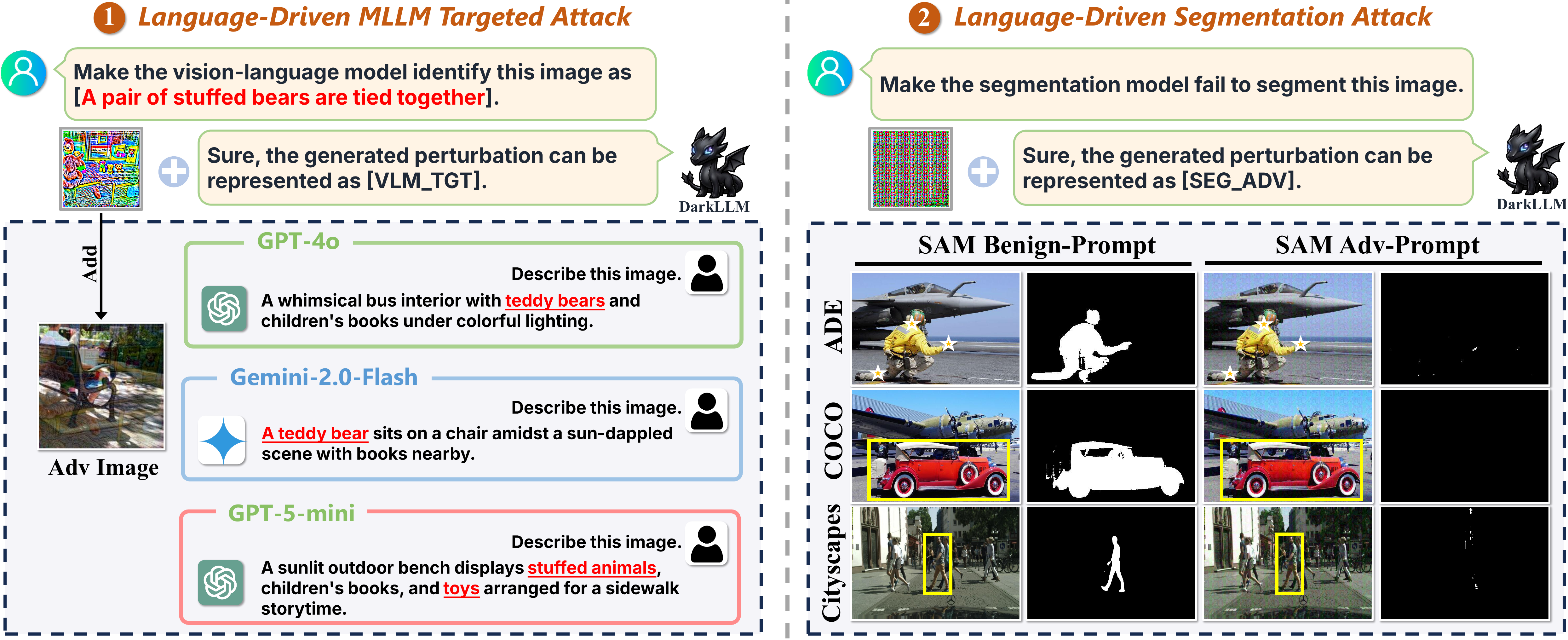}
\end{center}
\vspace{-1em}
    \caption{Visualization of DarkLLM for language-driven attacks.}
\label{fig:visualization}
\vspace{-1em}
\end{figure*}

\paragraph{Untargeted Attacks on Vision-Language Models.} 
We evaluate the non-targeted attack performance of DarkLLM across three vision-language tasks, with results summarized in Table~\ref{tab:untargeted_vlm}. Two key observations emerge. First, DarkLLM achieves strong and consistent performance across zero-shot classification and image captioning. It outperforms prior methods designed for both image classifiers (e.g., GD-UAP~\cite{mopuri2018generalizable}, TRM-UAP~\cite{liu2023trm}) and CLIP-based attacks (e.g., ETU~\cite{zhang2024universal}, C-PGC~\cite{fang2025one}), while remaining competitive with the strong baseline X-Transfer~\cite{huang2025x}. On image captioning, DarkLLM attains ASRs of 19.3\% on MSCOCO and 19.0\% on Flickr30k, slightly surpassing X-Transfer. Second, DarkLLM demonstrates clear advantages on image--text retrieval, achieving higher ASR than all baselines, including X-Transfer, indicating improved transferability across CLIP encoders. Collectively, these results demonstrate that DarkLLM achieves strong cross-task transferability, effectively generalizing across both discriminative and generative vision--language models.
\paragraph{Targeted Attacks on Frontier MLLMs.}
We evaluate the black-box transferability of DarkLLM across six frontier commercial MLLMs, with results summarized in Table~\ref{tab:attack_mllm}. As a universal method, DarkLLM consistently achieves strong performance compared to the previous state-of-the-art baseline, AnyAttack~\cite{zhang2025anyattack}, outperforming it on most evaluated models. Notably, DarkLLM achieves 67.7\% ASR on GPT-4o and 60.6\% on Gemini-3-Flash under the InternVL-2B setting, surpassing AnyAttack by clear margins. We further observe that scaling the controller model from InternVL-1B to InternVL-2B consistently improves both ASR and semantic similarity (AvgSim), suggesting that stronger language modeling capability enhances adversarial generation. While sample-wise methods such as M-Attack~\cite{li2025frustratingly} can achieve higher ASRs through per-image optimization, DarkLLM attains competitive results using a single universal perturbation, offering a more efficient and scalable alternative. Finally, as illustrated in Figure~\ref{fig:visualization} and Figure~\ref{fig:visualization_mllm}, DarkLLM is able to steer these models toward generating adversary-specified content via natural language instructions, highlighting the potential risks of language-guided adversarial generation in multimodal systems.
\definecolor{lightgray}{rgb}{0.90,0.90,0.90}   

\begin{table*}[!t]
\setlength{\abovecaptionskip}{4pt}
  \centering
    \caption{The mIoU (\%) results for promptable image segmentation task across different SAM models. The best results are \textbf{boldfaced}, and the second-best results are \underline{underlined}. \textcolor{red}{Red} numbers indicate the absolute mIoU drop caused by DarkLLM relative to the clean results.}
      \resizebox{\textwidth}{!}{
    \begin{tabular}{
    >{\centering\arraybackslash}m{1.1cm}|
    >{\raggedright\arraybackslash}m{3.0cm}|
    >{\raggedright\arraybackslash}m{1.00cm}
    >{\raggedright\arraybackslash}m{1.10cm}
    >{\raggedright\arraybackslash}m{1.10cm}|
    >{\raggedright\arraybackslash}m{1.10cm}
    >{\raggedright\arraybackslash}m{1.10cm}
    >{\raggedright\arraybackslash}m{1.10cm}| 
    >{\raggedright\arraybackslash}m{1.10cm}
    >{\raggedright\arraybackslash}m{1.10cm}
    >{\raggedright\arraybackslash}m{1.10cm}|
    >{\raggedright\arraybackslash}m{1.10cm}
    >{\raggedright\arraybackslash}m{1.10cm}
    >{\raggedright\arraybackslash}m{1.00cm}}
    \toprule[1.5pt]
    \multicolumn{2}{c|}{\textbf{Setting}} & \multicolumn{3}{c|}{\textbf{SAM-Base}~\cite{kirillov2023segment}} & \multicolumn{3}{c|}{\textbf{SAM-Huge}~\cite{kirillov2023segment}}       & \multicolumn{3}{c|}{\textbf{HQ-SAM-Huge}~\cite{ke2023segment}}    & \multicolumn{3}{c}{\textbf{SAM2-Hiera-Large}~\cite{ravi2024sam}} \\
 \cmidrule(lr){3-5} \cmidrule(lr){6-8} \cmidrule(lr){9-11} \cmidrule(lr){12-14}  \textbf{Prompt}  & \textbf{Method} & \texttt{\textbf{ADE}} & \texttt{\textbf{COCO}}  & \texttt{\textbf{CITY}}  & \texttt{\textbf{ADE}} & \texttt{\textbf{COCO}}  & \texttt{\textbf{CITY}}  & \texttt{\textbf{ADE}} & \texttt{\textbf{COCO}} & \texttt{\textbf{CITY}}  & \texttt{\textbf{ADE}} & \texttt{\textbf{COCO}}  & \texttt{\textbf{CITY}} \\
    \midrule[0.75pt]
    \multirow{6}[2]{*}{Point} & Clean & 50.3 & 44.9 & 38.0 & 53.6 & 48.5 & 39.7 & 52.4 & 47.3 & 32.0 & 57.2 & 55.2 & 41.7  \\
          & AttackSAM~\cite{zhang2023attack} & 46.8 & 42.0 & 24.6 & 51.7 & 47.5 & 32.2 & 51.0 & 46.5 & 26.8 & 55.9 & 55.0 & 41.4    \\
          & UAD~\cite{lu2024unsegment} & \underline{42.7} & \underline{39.4} & \underline{8.6} & 48.0 & \underline{44.6} & \textbf{16.0} & 46.5 & 43.6 & \underline{14.4} & 52.6 & 52.7 & \underline{26.0}    \\
          & DarkSAM-pt~\cite{zhou2024darksam}  & 43.2 & 41.6 & 29.6 & 53.1 & 48.7 & 38.4 & 50.1 & 46.7 & 30.6 & 52.5 & 49.1 & 38.5    \\
          & DarkSAM-bx~\cite{zhou2024darksam} & 46.1 & 43.3 & 32.6 & 52.7 & 48.7 & 38.4 & 48.4 & 45.2 & 28.2 & 52.5 & 48.7 & 39.2    \\
          & UAP-SAM2~\cite{zhou2025vanish} & 42.8 & 39.6 & 28.2 & \underline{47.2} & 44.9 & 32.9 & \underline{42.1} & \underline{40.6} & 24.9  & \underline{45.9} & \underline{44.1} & 29.2 \\
          & PATA~\cite{zheng2024black} & 44.0 & 40.1 & 13.0 & 48.7 & 45.0 & 23.3 & 47.5 & 44.3 & 21.2 & 53.0 & 52.7 & 31.5    \\
          & PATA++~\cite{zheng2024black} & 43.9 & 40.1 & 13.2 & 48.6 & 44.9 & 23.3 & 47.5 & 44.2 & 21.3 & 52.7 & 52.8 & 31.7    \\
          
          & \cellcolor{blue!10} DarkLLM (Ours) 
          & \cellcolor{blue!10} \textbf{2.9}\textbf{\textcolor{red}{\scriptsize {(47.4↓)}}}      
          & \cellcolor{blue!10} \textbf{3.5}\textbf{\textcolor{red}{\scriptsize {(41.4↓)}}}     
          & \cellcolor{blue!10} \textbf{0.4}\textbf{\textcolor{red}{\scriptsize {(37.6↓)}}}       
          & \cellcolor{blue!10} \textbf{21.5}\textbf{\textcolor{red}{\scriptsize {(32.1↓)}}}      
          & \cellcolor{blue!10} \textbf{24.3}\textbf{{\textcolor{red}{\scriptsize {(24.2↓)}}}}      
          & \cellcolor{blue!10} \underline{19.2}\textbf{{\textcolor{red}{\scriptsize {(20.5↓)}}} }   
          & \cellcolor{blue!10} \textbf{18.9}\textbf{\textcolor{red}{\scriptsize {(33.5↓)}}}      
          & \cellcolor{blue!10} \textbf{21.9}\textbf{\textcolor{red}{\scriptsize {(25.4↓)}}}      
          & \cellcolor{blue!10} \textbf{14.2}\textbf{{\textcolor{red}{\scriptsize {(17.8↓)}}}}  
          & \cellcolor{blue!10} \textbf{20.2}\textbf{\textcolor{red}{\scriptsize {(37.0↓)}}}  
          & \cellcolor{blue!10} \textbf{23.4}\textbf{\textcolor{red}{\scriptsize {(31.8↓)}}}  
          & \cellcolor{blue!10} \textbf{15.1}\textbf{\textcolor{red}{\scriptsize {(26.6↓)}}}  
          \\
    \midrule[0.75pt]
    \multirow{6}[2]{*}{Box } & Clean & 72.8 & 71.4 & 61.2 & 74.5 & 73.2 & 61.5 & 69.4 & 65.2 & 50.2 & 74.9 & 73.9 & 62.7   \\
          & AttackSAM~\cite{zhang2023attack} & 70.8 & 69.4 & 49.3 & 73.5 & 72.4 & 56.3 & 68.4 & 64.9 & 45.9 & 74.3 & 73.5 & 59.0    \\
          & UAD~\cite{lu2024unsegment} & 68.3 & 68.0 & \underline{39.7} & 71.3 & \underline{71.1} & \underline{47.9} & 65.8 & 63.7 & \underline{33.5} & 73.0 & 72.7 & \underline{53.4}    \\
          & DarkSAM-pt~\cite{zhou2024darksam}  & \underline{67.4} & \underline{67.5} & 51.3 & 73.7 & 72.9 & 58.8 & 67.6 & 64.4 & 46.1 & 73.9 & 73.2 & 59.5    \\
          & DarkSAM-bx~\cite{zhou2024darksam} & 68.9 & 68.5 & 53.8 & 73.6 & 72.9 & 59.1 & 66.1 & 63.4 & 44.2 & 74.1 & 73.2 & 60.0    \\
          & UAP-SAM2~\cite{zhou2025vanish} & 69.1 & 68.3 & 54.5 & \underline{71.2} & 71.2 & 55.9 & \underline{63.5} & \underline{61.9} & 42.2 & \underline{71.3} & \underline{71.2} & 55.4    \\
          & PATA~\cite{zheng2024black} & 68.8 & 68.1 & 43.1 & 71.7 & \underline{71.1} & 50.4 & 66.4 & 63.8 & 38.4 & 73.1 & 72.7 & 54.9    \\
          & PATA++~\cite{zheng2024black} & 68.7 & 68.0 & 43.5 & 71.6 & \underline{71.1} & 50.5 & 66.4 & 63.8 & 38.6 & 73.1 & 72.7 & 55.0    \\
          
          & \cellcolor{blue!10} DarkLLM (Ours) 
          & \cellcolor{blue!10} \textbf{27.4}\textbf{\textcolor{red}{\scriptsize {(45.4↓)}}}      
          & \cellcolor{blue!10} \textbf{30.5}\textbf{\textcolor{red}{\scriptsize {(40.9↓)}}}     
          & \cellcolor{blue!10} \textbf{30.2}\textbf{\textcolor{red}{\scriptsize {(31.0↓)}}}       
          & \cellcolor{blue!10} \textbf{46.9}\textbf{\textcolor{red}{\scriptsize {(27.6↓)}}}      
          & \cellcolor{blue!10} \textbf{55.1}\textbf{{\textcolor{red}{\scriptsize {(18.1↓)}}}   }   
          & \cellcolor{blue!10} \textbf{43.4}\textbf{{\textcolor{red}{\scriptsize {(18.1↓)}}}   } 
          & \cellcolor{blue!10} \textbf{40.9}\textbf{\textcolor{red}{\scriptsize {(28.5↓)}}}      
          & \cellcolor{blue!10} \textbf{47.1}\textbf{\textcolor{red}{\scriptsize {(18.1↓)}}}      
          & \cellcolor{blue!10} \textbf{32.3}\textbf{{\textcolor{red}{\scriptsize {(17.9↓)}}}}  
          & \cellcolor{blue!10} \textbf{49.5}\textbf{\textcolor{red}{\scriptsize {(25.4↓)}}}  
          & \cellcolor{blue!10} \textbf{55.7}\textbf{\textcolor{red}{\scriptsize {(18.2↓)}}}  
          & \cellcolor{blue!10} \textbf{38.8}\textbf{\textcolor{red}{\scriptsize {(23.9↓)}}}  
          \\
    \bottomrule[1.5pt]
    \end{tabular}%
    }
  \label{tab:attack_sam}%
    \vspace{-0.4cm}
\end{table*}%
\definecolor{lightgray}{rgb}{0.90,0.90,0.90}   

\begin{table*}[!t]
\setlength{\abovecaptionskip}{4pt}
  \centering
    \caption{Non-targeted and targeted ASR (\%) results of DarkLLM across different model families. The best results are \textbf{boldfaced}, and the second-best results are \underline{underlined}. \textcolor{red}{Red} numbers indicate the absolute ASR improvement of DarkLLM over domain-specific methods. }
      \resizebox{\textwidth}{!}{
    \begin{tabular}{
    >{\centering\arraybackslash}m{1.0cm}|
    >{\raggedright\arraybackslash}m{2.2cm}|
    >{\raggedright\arraybackslash}m{1.15cm}
    >{\raggedright\arraybackslash}m{1.15cm}
    >{\raggedright\arraybackslash}m{1.15cm}|
    >{\raggedright\arraybackslash}m{1.15cm}
    >{\raggedright\arraybackslash}m{1.15cm}
    >{\raggedright\arraybackslash}m{1.15cm}
    >{\raggedright\arraybackslash}m{1.15cm}|
    >{\raggedright\arraybackslash}m{1.15cm}
    >{\raggedright\arraybackslash}m{1.15cm}
    >{\raggedright\arraybackslash}m{1.15cm}
    >{\raggedright\arraybackslash}m{1.15cm}
    }
    \toprule[1.5pt]
    \multicolumn{2}{c|}{\textbf{Setting}} & \multicolumn{3}{c|}{\textbf{SAM2-Hiera-Large}}       & \multicolumn{4}{c|}{\textbf{CLIP (Avg@4)}}    & \multicolumn{4}{c}{\textbf{Commercial MLLMs}} \\
 \cmidrule(lr){3-5} \cmidrule(lr){6-9} \cmidrule(lr){10-13}   \textbf{Prompt}  & \textbf{Method}   & \texttt{\textbf{ADE}} & \texttt{\textbf{COCO}}  & \texttt{\textbf{CITY}}  & \texttt{\textbf{Food}} & \texttt{\textbf{Image}}  & \texttt{\textbf{Cars}} & \texttt{\textbf{STL}} & \texttt{\textbf{QwenVL}} & \texttt{\textbf{GPT-4o}} & \texttt{\textbf{GPT-5}} & \texttt{\textbf{Gemini}} \\
    \midrule[0.75pt]
    \multirow{3}[2]{*}{Box} 
          & DarkSAM~\cite{zhou2024darksam} & \cellcolor{gray!20} 2.0 & \cellcolor{gray!20}1.4 & \cellcolor{gray!20}\underline{6.5} & 5.4 & 4.1 & 5.0 & 0.9 & -- & -- & -- & -- \\
       & X-Transfer~\cite{huang2025x} & \underline{3.3} & \underline{2.2} & 5.1 & \cellcolor{gray!20}\textbf{83.6} & \cellcolor{gray!20}\textbf{63.4} & \cellcolor{gray!20}\textbf{69.7} & \cellcolor{gray!20}\textbf{49.3}  & -- & -- & -- & -- \\
          & AnyAttack~\cite{zhang2025anyattack} & 2.6 & 1.7 & 4.1 & 50.4 & 35.1 & 31.1 & 27.9 & 
          \cellcolor{gray!20} \textbf{56.0} & \cellcolor{gray!20}\underline{60.0} & 
          \cellcolor{gray!20}\underline{53.0} & 
          \cellcolor{gray!20}\underline{46.0} \\
          & \cellcolor{blue!5} DarkLLM 
          & \cellcolor{blue!5} \textbf{24.8}\textbf{\textcolor{red}{\scriptsize {(22.8↑)}}}
          & \cellcolor{blue!5} \textbf{19.2}\textbf{\textcolor{red}{\scriptsize {(17.8↑)}}}
          & \cellcolor{blue!5} \textbf{30.9}\textbf{\textcolor{red}{\scriptsize {(24.4↑)}}}
          & \cellcolor{blue!5} \underline{68.8}\textbf{\textcolor{red}{\scriptsize {(14.8↓)}}}
          & \cellcolor{blue!5} \underline{50.5}\textbf{\textcolor{red}{\scriptsize {(12.9↓)}}}
          & \cellcolor{blue!5} \underline{50.1}\textbf{\textcolor{red}{\scriptsize {(19.6↓)}}}
          & \cellcolor{blue!5} \underline{33.1}\textbf{\textcolor{red}{\scriptsize {(16.2↓)}}}
          & \cellcolor{blue!5} \underline{49.5}\textbf{\textcolor{red}{\scriptsize {(6.5↓)}}}
          & \cellcolor{blue!5} \textbf{64.6}\textbf{\textcolor{red}{\scriptsize {(4.6↑)}}}
          & \cellcolor{blue!5} \textbf{54.5}\textbf{\textcolor{red}{\scriptsize {(1.5↑)}}}
          & \cellcolor{blue!5} \textbf{52.5}\textbf{\textcolor{red}{\scriptsize {(6.5↑)}}}
          \\
    \bottomrule[1.5pt]
    \end{tabular}%
    }
  \label{tab:attack_transferability}%
    \vspace{-0.2cm}
\end{table*}%
\begin{figure*}[!t] 
\setlength{\abovecaptionskip}{4pt}
  \centering
     \subcaptionbox{Number of Prompts}
{\vspace{2pt}\includegraphics[width=0.24\textwidth]{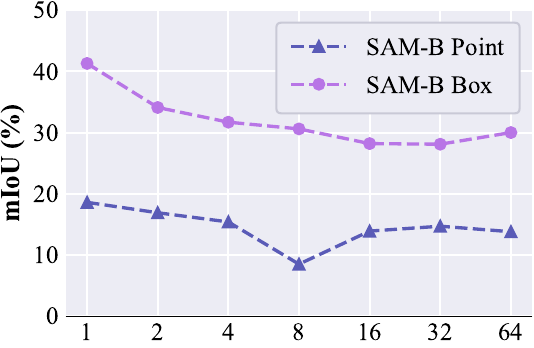}}
       \subcaptionbox{Prompt Type}
{\vspace{2pt}\includegraphics[width=0.24\textwidth]{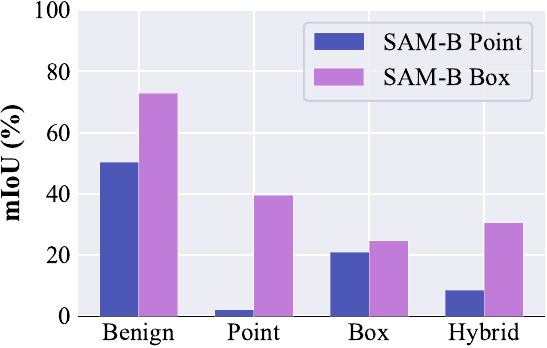}} 
        \subcaptionbox{Loss Function Type}
{\vspace{2pt}\includegraphics[width=0.24\textwidth]{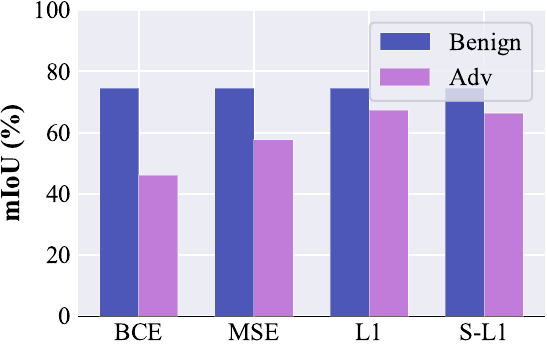}} 
       \subcaptionbox{Surrogate Model}
{\vspace{2pt}\includegraphics[width=0.24\textwidth]{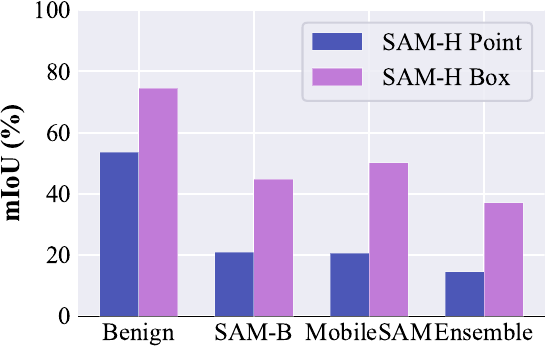}} 
        \subcaptionbox{Targeted Surrogate}{\includegraphics[width=0.24\textwidth]{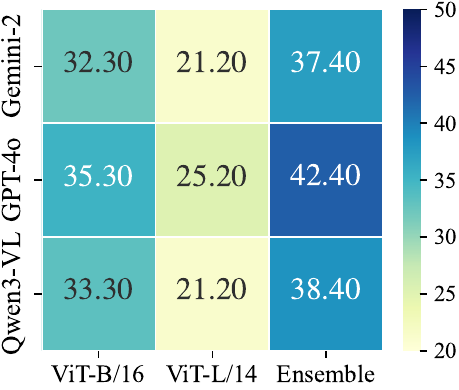}}
    \subcaptionbox{Untargeted Surrogate}{\includegraphics[width=0.24\textwidth]{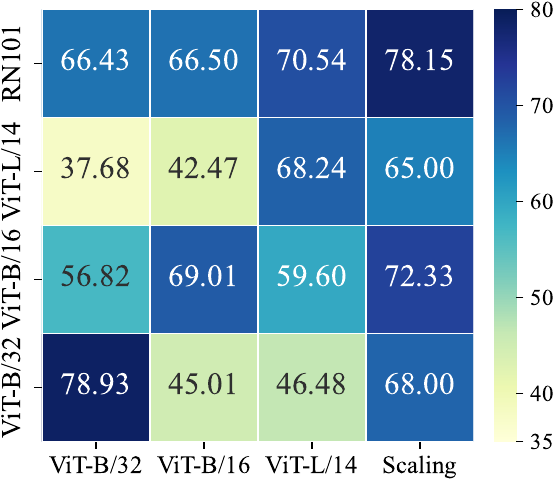}}
     \subcaptionbox{Training Strategy}
     {\includegraphics[width=0.24\textwidth]{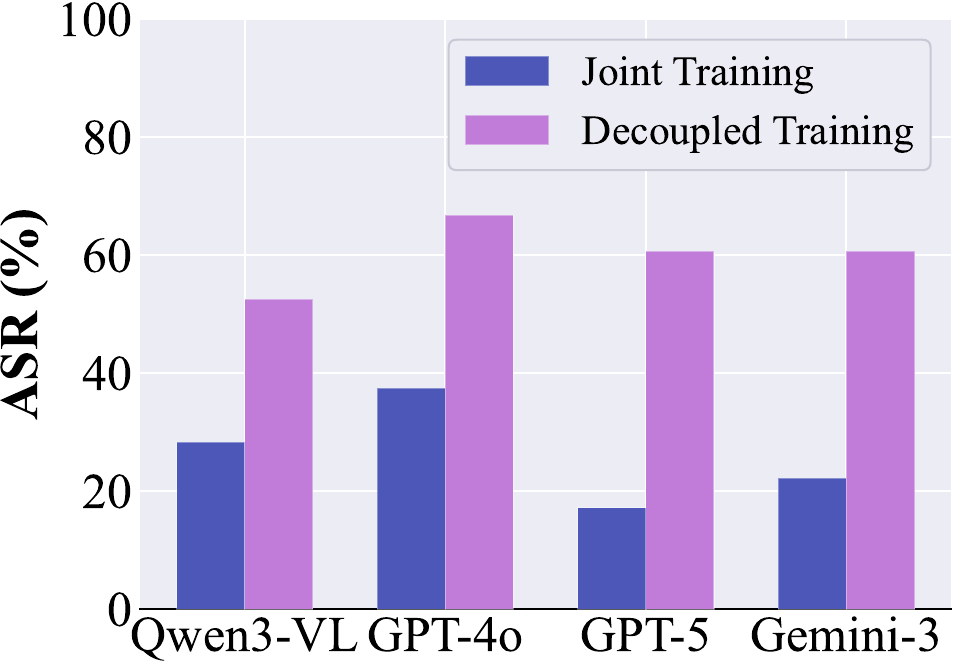}}
     \subcaptionbox{Training Data Scale}
     {\includegraphics[width=0.24\textwidth]{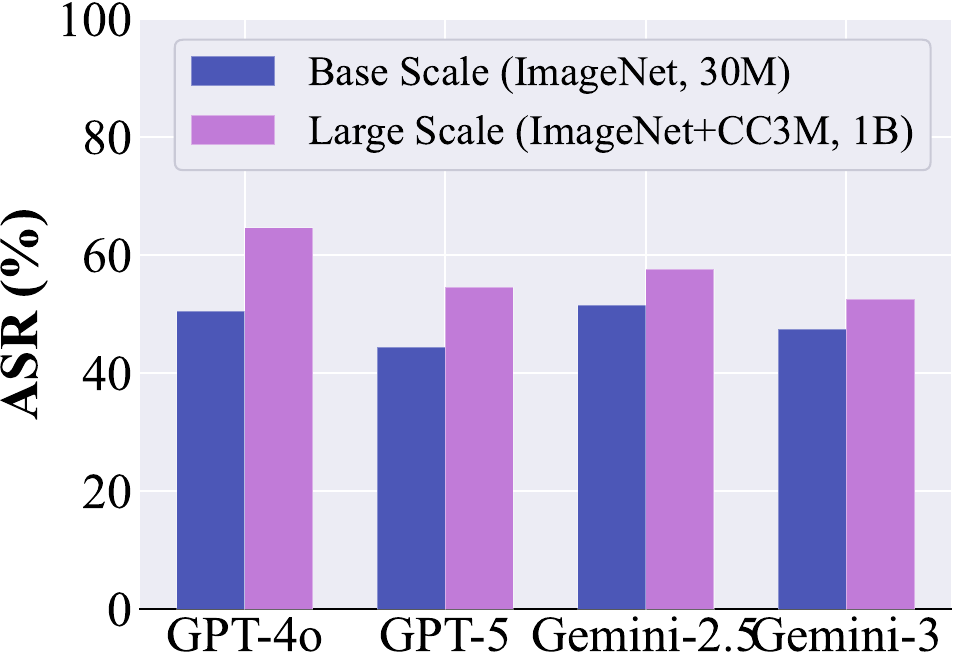}}
      \caption{Ablation study on the effects of different factors on the attack performance of DarkLLM.}
       \label{fig:ablation}
       \vspace{-0.7cm}
\end{figure*}

\paragraph{Untargeted Attacks on Segmentation Models.}
We investigate the effectiveness of DarkLLM in disrupting promptable segmentation models across four SAM variants and three datasets. Table~\ref{tab:attack_sam} reveals three key observations.
\textbf{(1)} DarkLLM consistently induces substantial mIoU degradation under both point and box prompts, demonstrating robustness to different interaction modalities.
\textbf{(2)} The performance drop remains significant across larger and more advanced models, including SAM-Huge, HQ-SAM-Huge, and SAM2-Large, indicating strong cross-backbone transferability where many baselines show limited effectiveness. 
\textbf{(3)} DarkLLM achieves larger performance degradation than the strongest baseline UAD~\cite{lu2024unsegment} across most settings, establishing consistently strong attack effectiveness. Figure~\ref{fig:visualization}, Figure~\ref{fig:visualization_sam_b} --\ref{fig:visualization_sam_h} illustrate that DarkLLM effectively disrupts mask generation across different SAM models. Results in Table~\ref{tab:appendix_sam} --\ref{tab:appendix_sam2} further confirm the superiority of our method.
\vspace{-8pt}
\paragraph{Unified Attack Scalability.}
We assess the scalability of DarkLLM across heterogeneous models, tasks, and domains. Table~\ref{tab:attack_transferability} reveals two key findings.
\textbf{(1)} DarkLLM achieves strong and consistent performance across all tasks. It induces substantial degradation on SAM2 across ADE, COCO, and CITY, while simultaneously attaining high ASRs on CLIP and commercial MLLMs. Compared to prior methods, DarkLLM outperforms DarkSAM on segmentation and achieves competitive or superior performance to X-Transfer and AnyAttack across CLIP and MLLMs.
\textbf{(2)} More importantly, Table~\ref{tab:attack_transferability} highlights a key limitation of model-specific attacks: methods such as X-Transfer, tailored for CLIP, exhibit limited effectiveness on SAM2. In contrast, DarkLLM generates a \textit{single} perturbation that simultaneously degrades SAM2 and maintains strong transferability to CLIP and MLLMs, revealing shared vulnerabilities across diverse foundation model architectures.

\begin{wrapfigure}[10]{R}{0.5\textwidth}
\setlength{\abovecaptionskip}{4pt}
\vspace{-10pt}

\centering

\subcaptionbox{SAM}
{\includegraphics[width=0.49\linewidth]{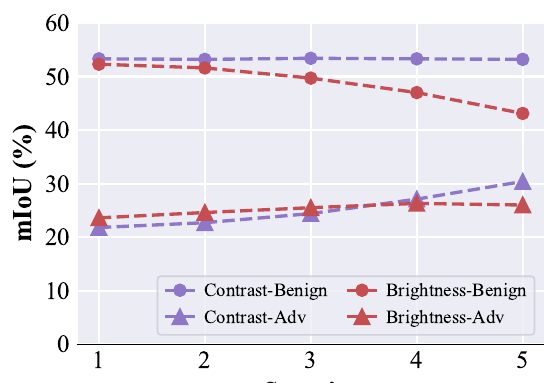}}
\hfill
\subcaptionbox{CLIP}
{\includegraphics[width=0.49\linewidth]{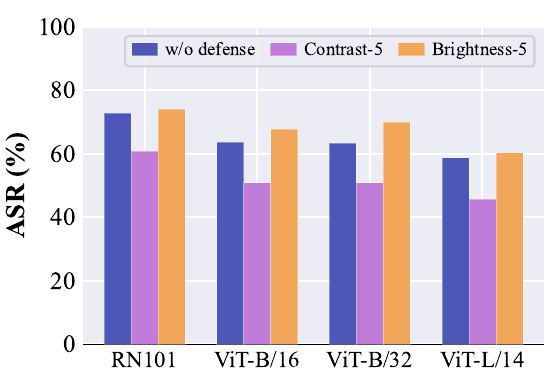}}

\caption{Attack effectiveness of DarkLLM under defense mechanisms.}
\label{fig:ablation_defense}

\vspace{-10pt}
\end{wrapfigure}

\subsection{Method Analysis}
In this section, we investigate the different factors on the attack performance of DarkLLM.

\noindent \textbf{Attack Optimization on SAM.}\quad
We analyze several factors that improve the attack effectiveness and transferability on SAM from different perspectives. As shown in Figure~\ref{fig:ablation}(a), increasing the number of prompts used during optimization consistently strengthens the attack, leading to lower mIoU, with $N{=}8$ achieving a strong trade-off between performance and efficiency. Figure~\ref{fig:ablation}(b) further shows that attacks optimized with a single prompt type (point or box) generalize poorly across prompt formats, while a hybrid optimization strategy that jointly leverages both prompts yields the most robust performance. In addition, Figure~\ref{fig:ablation}(c) evaluates the impact of different loss functions on cross-backbone transfer. BCE demonstrates significantly stronger transferability to SAM-Huge compared to MSE, $\ell_1$, and Smooth-$\ell_1$, and is therefore adopted as our default choice. 

\noindent \textbf{Surrogate Model Selection.}\quad
We analyze the impact of surrogate model selection on attack transferability. As shown in Figure~\ref{fig:ablation}(d), using a single SAM-Base surrogate enables effective transfer to SAM-Huge, while combining SAM-Base with MobileSAM further strengthens the attack. For targeted attacks, we optimize DarkLLM using CLIP surrogate models following~\cite{zhao2023evaluating,zhang2025anyattack}. As shown in Figure~\ref{fig:ablation}(e), employing a surrogate ensemble (ViT-B/16 + ViT-L/14) consistently improves targeted ASR across different MLLMs compared to using a single surrogate. For untargeted attacks, Figure~\ref{fig:ablation}(f) shows that the surrogate scaling strategy, which leverages a diverse pool of CLIP backbones, achieves the best transferability and consistently outperforms any single fixed surrogate. These results indicate that increasing surrogate diversity is a key factor in enhancing both targeted and untargeted attack performance, consistent with prior findings~\cite{huang2025x,li2025frustratingly,zhang2025anyattack}.

\noindent \textbf{Training Strategy and Data Scaling.}\quad
We further analyze training-related factors that affect the targeted attack performance of DarkLLM on MLLMs. Figure~\ref{fig:ablation}(g) shows that our decoupled progressive training consistently achieves higher ASR than joint training across different MLLMs. This suggests that separating heterogeneous attack objectives during training helps reduce optimization interference and better preserve the instruction-sensitive targeted attack capability. We also study the effect of training data scale. Specifically, we compare two training configurations with different effective data scales: ImageNet training for 24 epochs (approximately 30M effective samples) and large-scale training on ImageNet+CC3M with adjusted epochs (approximately 1B effective samples). As shown in Figure~\ref{fig:ablation}(h), increasing the effective training scale consistently improves targeted ASR across different MLLMs, indicating that larger-scale training enhances the transferability and generalization.

\noindent \textbf{Attack Effectiveness Under Defenses.}\quad
We evaluate the effectiveness of DarkLLM against potential preprocessing-based defenses. Specifically, we consider two widely used image corruption methods, contrast and brightness, as defensive transformations. For SAM, we conduct experiments on the ADE20K dataset using point prompts on SAM-Huge, where the defense strength is controlled by five severity levels (from 1 to 5). As shown in Figure~\ref{fig:ablation_defense}(a), our method consistently degrades the segmentation performance, maintaining mIoU well below the clean baseline across all severity levels. For CLIP, we apply the strongest corruption (severity 5) and evaluate zero-shot classification on ImageNet across different backbones. As shown in Figure~\ref{fig:ablation_defense}(b), DarkLLM achieves high ASR under both defenses, with only a moderate drop under contrast, demonstrating strong robustness against such preprocessing-based defenses.

\section{Conclusion}
\label{sec:conclusion}
In this work, we present DarkLLM, a language-guided adversarial attack framework that leverages large language models as dynamic controllers for visual adversarial generation. By bridging natural-language instructions with pixel-level perturbations, DarkLLM generalizes beyond traditional optimization-based attacks and unifies targeted, untargeted, segmentation, and multi-model attacks within a single framework. Comprehensive evaluations demonstrate that our approach achieves strong transferability across diverse vision foundation models. We hope this work provides useful insights for the community in understanding and developing both LLM-based attacks and defenses.
{
    \small
    \bibliographystyle{plain}
    \bibliography{main}

@article{sun2026sama,
  title={Sama: Towards multi-turn referential grounded video chat with large language models},
  author={Sun, Ye and Zhang, Hao and Ding, Henghui and Zhang, Tiehua and Ma, Xingjun and Jiang, Yu-Gang},
  journal={Advances in Neural Information Processing Systems},
  volume={38},
  pages={47065--47091},
  year={2026}
}

@article{chen2024expanding,
  title={Expanding performance boundaries of open-source multimodal models with model, data, and test-time scaling},
  author={Chen, Zhe and Wang, Weiyun and Cao, Yue and Liu, Yangzhou and Gao, Zhangwei and Cui, Erfei and Zhu, Jinguo and Ye, Shenglong and Tian, Hao and Liu, Zhaoyang and others},
  journal={arXiv preprint arXiv:2412.05271},
  year={2024}
}

@inproceedings{radford2021learning,
  title={Learning transferable visual models from natural language supervision},
  author={Radford, Alec and Kim, Jong Wook and Hallacy, Chris and Ramesh, Aditya and Goh, Gabriel and Agarwal, Sandhini and Sastry, Girish and Askell, Amanda and Mishkin, Pamela and Clark, Jack and others},
  booktitle={International conference on machine learning},
  pages={8748--8763},
  year={2021},
  organization={PmLR}
}

@inproceedings{li2023blip,
  title={Blip-2: Bootstrapping language-image pre-training with frozen image encoders and large language models},
  author={Li, Junnan and Li, Dongxu and Savarese, Silvio and Hoi, Steven},
  booktitle={International conference on machine learning},
  pages={19730--19742},
  year={2023},
  organization={PMLR}
}

@article{liu2023visual,
  title={Visual instruction tuning},
  author={Liu, Haotian and Li, Chunyuan and Wu, Qingyang and Lee, Yong Jae},
  journal={Advances in neural information processing systems},
  volume={36},
  pages={34892--34916},
  year={2023}
}

@inproceedings{kirillov2023segment,
  title={Segment anything},
  author={Kirillov, Alexander and Mintun, Eric and Ravi, Nikhila and Mao, Hanzi and Rolland, Chloe and Gustafson, Laura and Xiao, Tete and Whitehead, Spencer and Berg, Alexander C and Lo, Wan-Yen and others},
  booktitle={Proceedings of the IEEE/CVF international conference on computer vision},
  pages={4015--4026},
  year={2023}
}

@article{achiam2023gpt,
  title={Gpt-4 technical report},
  author={Achiam, Josh and Adler, Steven and Agarwal, Sandhini and Ahmad, Lama and Akkaya, Ilge and Aleman, Florencia Leoni and Almeida, Diogo and Altenschmidt, Janko and Altman, Sam and Anadkat, Shyamal and others},
  journal={arXiv preprint arXiv:2303.08774},
  year={2023}
}

@article{team2023gemini,
  title={Gemini: a family of highly capable multimodal models},
  author={Team, Gemini and Anil, Rohan and Borgeaud, Sebastian and Alayrac, Jean-Baptiste and Yu, Jiahui and Soricut, Radu and Schalkwyk, Johan and Dai, Andrew M and Hauth, Anja and Millican, Katie and others},
  journal={arXiv preprint arXiv:2312.11805},
  year={2023}
}

@article{szegedy2013intriguing,
  title={Intriguing properties of neural networks},
  author={Szegedy, Christian and Zaremba, Wojciech and Sutskever, Ilya and Bruna, Joan and Erhan, Dumitru and Goodfellow, Ian and Fergus, Rob},
  journal={arXiv preprint arXiv:1312.6199},
  year={2013}
}

@article{goodfellow2014explaining,
  title={Explaining and harnessing adversarial examples},
  author={Goodfellow, Ian J and Shlens, Jonathon and Szegedy, Christian},
  journal={arXiv preprint arXiv:1412.6572},
  year={2014}
}

@article{madry2017towards,
  title={Towards deep learning models resistant to adversarial attacks},
  author={Madry, Aleksander and Makelov, Aleksandar and Schmidt, Ludwig and Tsipras, Dimitris and Vladu, Adrian},
  journal={arXiv preprint arXiv:1706.06083},
  year={2017}
}

@article{ma2026safety,
  title={Safety at scale: A comprehensive survey of large model and agent safety},
  author={Ma, Xingjun and Gao, Yifeng and Wang, Yixu and Wang, Ruofan and Wang, Xin and Sun, Ye and Ding, Yifan and Xu, Hengyuan and Chen, Yunhao and Zhao, Yunhan and others},
  journal={Foundations and Trends in Privacy and Security},
  volume={8},
  number={3-4},
  pages={1--240},
  year={2026},
  publisher={Emerald Publishing Limited}
}

@article{brown2020language,
  title={Language models are few-shot learners},
  author={Brown, Tom and Mann, Benjamin and Ryder, Nick and Subbiah, Melanie and Kaplan, Jared D and Dhariwal, Prafulla and Neelakantan, Arvind and Shyam, Pranav and Sastry, Girish and Askell, Amanda and others},
  journal={Advances in neural information processing systems},
  volume={33},
  pages={1877--1901},
  year={2020}
}

@article{tian2024visual,
  title={Visual autoregressive modeling: Scalable image generation via next-scale prediction},
  author={Tian, Keyu and Jiang, Yi and Yuan, Zehuan and Peng, Bingyue and Wang, Liwei},
  journal={Advances in neural information processing systems},
  volume={37},
  pages={84839--84865},
  year={2024}
}

@article{singhal2025toward,
  title={Toward expert-level medical question answering with large language models},
  author={Singhal, Karan and Tu, Tao and Gottweis, Juraj and Sayres, Rory and Wulczyn, Ellery and Amin, Mohamed and Hou, Le and Clark, Kevin and Pfohl, Stephen R and Cole-Lewis, Heather and others},
  journal={Nature Medicine},
  volume={31},
  number={3},
  pages={943--950},
  year={2025},
  publisher={Nature Publishing Group US New York}
}

@article{xie2024show,
  title={Show-o: One single transformer to unify multimodal understanding and generation},
  author={Xie, Jinheng and Mao, Weijia and Bai, Zechen and Zhang, David Junhao and Wang, Weihao and Lin, Kevin Qinghong and Gu, Yuchao and Chen, Zhijie and Yang, Zhenheng and Shou, Mike Zheng},
  journal={arXiv preprint arXiv:2408.12528},
  year={2024}
}

@article{liu2023autodan,
  title={Autodan: Generating stealthy jailbreak prompts on aligned large language models},
  author={Liu, Xiaogeng and Xu, Nan and Chen, Muhao and Xiao, Chaowei},
  journal={arXiv preprint arXiv:2310.04451},
  year={2023}
}

@article{li2025autobackdoor,
  title={AutoBackdoor: Automating Backdoor Attacks via LLM Agents},
  author={Li, Yige and Li, Zhe and Zhao, Wei and Min, Nay Myat and Huang, Hanxun and Ma, Xingjun and Sun, Jun},
  journal={arXiv preprint arXiv:2511.16709},
  year={2025}
}

@article{ying2026safebench,
  title={Safebench: A safety evaluation framework for multimodal large language models},
  author={Ying, Zonghao and Liu, Aishan and Liang, Siyuan and Huang, Lei and Guo, Jinyang and Zhou, Wenbo and Liu, Xianglong and Tao, Dacheng},
  journal={International Journal of Computer Vision},
  volume={134},
  number={1},
  pages={18},
  year={2026},
  publisher={Springer}
}

@article{goodfellow2014generative,
  title={Generative adversarial nets},
  author={Goodfellow, Ian J and Pouget-Abadie, Jean and Mirza, Mehdi and Xu, Bing and Warde-Farley, David and Ozair, Sherjil and Courville, Aaron and Bengio, Yoshua},
  journal={Advances in neural information processing systems},
  volume={27},
  year={2014}
}

@inproceedings{moosavi2017universal,
  title={Universal adversarial perturbations},
  author={Moosavi-Dezfooli, Seyed-Mohsen and Fawzi, Alhussein and Fawzi, Omar and Frossard, Pascal},
  booktitle={Proceedings of the IEEE conference on computer vision and pattern recognition},
  pages={1765--1773},
  year={2017}
}

@article{zhao2023evaluating,
  title={On evaluating adversarial robustness of large vision-language models},
  author={Zhao, Yunqing and Pang, Tianyu and Du, Chao and Yang, Xiao and Li, Chongxuan and Cheung, Ngai-Man Man and Lin, Min},
  journal={Advances in Neural Information Processing Systems},
  volume={36},
  pages={54111--54138},
  year={2023}
}

@article{li2025frustratingly,
  title={A frustratingly simple yet highly effective attack baseline: Over 90\% success rate against the strong black-box models of gpt-4.5/4o/o1},
  author={Li, Zhaoyi and Zhao, Xiaohan and Wu, Dong-Dong and Cui, Jiacheng and Shen, Zhiqiang},
  journal={arXiv preprint arXiv:2503.10635},
  year={2025}
}

@article{jia2025adversarial,
  title={Adversarial Attacks against Closed-Source MLLMs via Feature Optimal Alignment},
  author={Jia, Xiaojun and Gao, Sensen and Qin, Simeng and Pang, Tianyu and Du, Chao and Huang, Yihao and Li, Xinfeng and Li, Yiming and Li, Bo and Liu, Yang},
  journal={arXiv preprint arXiv:2505.21494},
  year={2025}
}

@article{zhou2024darksam,
  title={Darksam: Fooling segment anything model to segment nothing},
  author={Zhou, Ziqi and Song, Yufei and Li, Minghui and Hu, Shengshan and Wang, Xianlong and Zhang, Leo Yu and Yao, Dezhong and Jin, Hai},
  journal={Advances in Neural Information Processing Systems},
  volume={37},
  pages={49859--49880},
  year={2024}
}

@inproceedings{zhang2022towards,
  title={Towards adversarial attack on vision-language pre-training models},
  author={Zhang, Jiaming and Yi, Qi and Sang, Jitao},
  booktitle={Proceedings of the 30th ACM International Conference on Multimedia},
  pages={5005--5013},
  year={2022}
}

@inproceedings{zhang2025anyattack,
  title={AnyAttack: Towards Large-scale Self-supervised Adversarial Attacks on Vision-language Models},
  author={Zhang, Jiaming and Ye, Junhong and Ma, Xingjun and Li, Yige and Yang, Yunfan and Chen, Yunhao and Sang, Jitao and Yeung, Dit-Yan},
  booktitle={Proceedings of the Computer Vision and Pattern Recognition Conference},
  pages={19900--19909},
  year={2025}
}

@article{xiao2018generating,
  title={Generating adversarial examples with adversarial networks},
  author={Xiao, Chaowei and Li, Bo and Zhu, Jun-Yan and He, Warren and Liu, Mingyan and Song, Dawn},
  journal={arXiv preprint arXiv:1801.02610},
  year={2018}
}

@inproceedings{ilyas2018black,
  title={Black-box adversarial attacks with limited queries and information},
  author={Ilyas, Andrew and Engstrom, Logan and Athalye, Anish and Lin, Jessy},
  booktitle={International conference on machine learning},
  pages={2137--2146},
  year={2018},
  organization={PMLR}
}

@inproceedings{andriushchenko2020square,
  title={Square attack: a query-efficient black-box adversarial attack via random search},
  author={Andriushchenko, Maksym and Croce, Francesco and Flammarion, Nicolas and Hein, Matthias},
  booktitle={European conference on computer vision},
  pages={484--501},
  year={2020},
  organization={Springer}
}

@inproceedings{dong2018boosting,
  title={Boosting adversarial attacks with momentum},
  author={Dong, Yinpeng and Liao, Fangzhou and Pang, Tianyu and Su, Hang and Zhu, Jun and Hu, Xiaolin and Li, Jianguo},
  booktitle={Proceedings of the IEEE conference on computer vision and pattern recognition},
  pages={9185--9193},
  year={2018}
}

@inproceedings{xie2019improving,
  title={Improving transferability of adversarial examples with input diversity},
  author={Xie, Cihang and Zhang, Zhishuai and Zhou, Yuyin and Bai, Song and Wang, Jianyu and Ren, Zhou and Yuille, Alan L},
  booktitle={Proceedings of the IEEE/CVF conference on computer vision and pattern recognition},
  pages={2730--2739},
  year={2019}
}

@article{huang2025x,
  title={X-Transfer Attacks: Towards Super Transferable Adversarial Attacks on CLIP},
  author={Huang, Hanxun and Erfani, Sarah and Li, Yige and Ma, Xingjun and Bailey, James},
  journal={arXiv preprint arXiv:2505.05528},
  year={2025}
}

@inproceedings{zhou2023advclip,
  title={Advclip: Downstream-agnostic adversarial examples in multimodal contrastive learning},
  author={Zhou, Ziqi and Hu, Shengshan and Li, Minghui and Zhang, Hangtao and Zhang, Yechao and Jin, Hai},
  booktitle={Proceedings of the 31st ACM International Conference on Multimedia},
  pages={6311--6320},
  year={2023}
}

@inproceedings{fang2025one,
  title={One perturbation is enough: On generating universal adversarial perturbations against vision-language pre-training models},
  author={Fang, Hao and Kong, Jiawei and Yu, Wenbo and Chen, Bin and Li, Jiawei and Wu, Hao and Xia, Shu-Tao and Xu, Ke},
  booktitle={Proceedings of the IEEE/CVF International Conference on Computer Vision},
  pages={4090--4100},
  year={2025}
}

@article{zhang2023attack,
  title={Attack-sam: Towards evaluating adversarial robustness of segment anything model},
  author={Zhang, Chenshuang and Zhang, Chaoning and Kang, Taegoo and Kim, Donghun and Bae, Sung-Ho and Kweon, In So},
  journal={arXiv preprint arXiv:2305.00866},
  volume={1},
  number={3},
  pages={5},
  year={2023}
}

@article{zhou2025vanish,
  title={Vanish into thin air: Cross-prompt universal adversarial attacks for sam2},
  author={Zhou, Ziqi and Hu, Yifan and Song, Yufei and Li, Zijing and Hu, Shengshan and Zhang, Leo Yu and Yao, Dezhong and Zheng, Long and Jin, Hai},
  journal={arXiv preprint arXiv:2510.24195},
  year={2025}
}

@article{luo2024jailbreakv,
  title={Jailbreakv: A benchmark for assessing the robustness of multimodal large language models against jailbreak attacks},
  author={Luo, Weidi and Ma, Siyuan and Liu, Xiaogeng and Guo, Xiaoyu and Xiao, Chaowei},
  journal={arXiv preprint arXiv:2404.03027},
  year={2024}
}

@article{zhang2023faster,
  title={Faster segment anything: Towards lightweight sam for mobile applications},
  author={Zhang, Chaoning and Han, Dongshen and Qiao, Yu and Kim, Jung Uk and Bae, Sung-Ho and Lee, Seungkyu and Hong, Choong Seon},
  journal={arXiv preprint arXiv:2306.14289},
  year={2023}
}

@inproceedings{deng2009imagenet,
  title={Imagenet: A large-scale hierarchical image database},
  author={Deng, Jia and Dong, Wei and Socher, Richard and Li, Li-Jia and Li, Kai and Fei-Fei, Li},
  booktitle={2009 IEEE conference on computer vision and pattern recognition},
  pages={248--255},
  year={2009},
  organization={Ieee}
}

@misc{2023xtuner,
    title={XTuner: A Toolkit for Efficiently Fine-tuning LLM},
    author={XTuner Contributors},
    howpublished = {\url{https://github.com/InternLM/xtuner}},
    year={2023}
}

@inproceedings{sharma2018conceptual,
  title={Conceptual captions: A cleaned, hypernymed, image alt-text dataset for automatic image captioning},
  author={Sharma, Piyush and Ding, Nan and Goodman, Sebastian and Soricut, Radu},
  booktitle={Proceedings of the 56th Annual Meeting of the Association for Computational Linguistics (Volume 1: Long Papers)},
  pages={2556--2565},
  year={2018}
}

@article{hu2022lora,
  title={Lora: Low-rank adaptation of large language models.},
  author={Hu, Edward J and Shen, Yelong and Wallis, Phillip and Allen-Zhu, Zeyuan and Li, Yuanzhi and Wang, Shean and Wang, Lu and Chen, Weizhu and others},
  journal={ICLR},
  volume={1},
  number={2},
  pages={3},
  year={2022}
}

@inproceedings{zhang2024universal,
  title={Universal adversarial perturbations for vision-language pre-trained models},
  author={Zhang, Peng-Fei and Huang, Zi and Bai, Guangdong},
  booktitle={Proceedings of the 47th International ACM SIGIR Conference on Research and Development in Information Retrieval},
  pages={862--871},
  year={2024}
}

@article{weng2024learning,
  title={Learning transferable targeted universal adversarial perturbations by sequential meta-learning},
  author={Weng, Juanjuan and Luo, Zhiming and Lin, Dazhen and Li, Shaozi},
  journal={Computers \& Security},
  volume={137},
  pages={103584},
  year={2024},
  publisher={Elsevier}
}

@inproceedings{liu2023trm,
  title={Trm-uap: Enhancing the transferability of data-free universal adversarial perturbation via truncated ratio maximization},
  author={Liu, Yiran and Feng, Xin and Wang, Yunlong and Yang, Wu and Ming, Di},
  booktitle={Proceedings of the IEEE/CVF International Conference on Computer Vision},
  pages={4762--4771},
  year={2023}
}

@article{mopuri2018generalizable,
  title={Generalizable data-free objective for crafting universal adversarial perturbations},
  author={Mopuri, Konda Reddy and Ganeshan, Aditya and Babu, R Venkatesh},
  journal={IEEE transactions on pattern analysis and machine intelligence},
  volume={41},
  number={10},
  pages={2452--2465},
  year={2018},
  publisher={IEEE}
}

@inproceedings{lu2024unsegment,
  title={Unsegment anything by simulating deformation},
  author={Lu, Jiahao and Yang, Xingyi and Wang, Xinchao},
  booktitle={Proceedings of the IEEE/CVF Conference on Computer Vision and Pattern Recognition},
  pages={24294--24304},
  year={2024}
}

@article{krizhevsky2009learning,
  title={Learning multiple layers of features from tiny images},
  author={Krizhevsky, Alex and Hinton, Geoffrey and others},
  year={2009},
  publisher={Toronto, ON, Canada}
}

@inproceedings{bossard2014food,
  title={Food-101--mining discriminative components with random forests},
  author={Bossard, Lukas and Guillaumin, Matthieu and Van Gool, Luc},
  booktitle={European conference on computer vision},
  pages={446--461},
  year={2014},
  organization={Springer}
}

@article{stallkamp2012man,
  title={Man vs. computer: Benchmarking machine learning algorithms for traffic sign recognition},
  author={Stallkamp, Johannes and Schlipsing, Marc and Salmen, Jan and Igel, Christian},
  journal={Neural networks},
  volume={32},
  pages={323--332},
  year={2012},
  publisher={Elsevier}
}

@article{krause2013collecting,
  title={Collecting a large-scale dataset of fine-grained cars},
  author={Krause, Jonathan and Deng, Jia and Stark, Michael and Fei-Fei, Li},
  year={2013}
}

@inproceedings{coates2011analysis,
  title={An analysis of single-layer networks in unsupervised feature learning},
  author={Coates, Adam and Ng, Andrew and Lee, Honglak},
  booktitle={Proceedings of the fourteenth international conference on artificial intelligence and statistics},
  pages={215--223},
  year={2011},
  organization={JMLR Workshop and Conference Proceedings}
}

@inproceedings{karpathy2015deep,
  title={Deep visual-semantic alignments for generating image descriptions},
  author={Karpathy, Andrej and Fei-Fei, Li},
  booktitle={Proceedings of the IEEE conference on computer vision and pattern recognition},
  pages={3128--3137},
  year={2015}
}

@inproceedings{lin2014microsoft,
  title={Microsoft coco: Common objects in context},
  author={Lin, Tsung-Yi and Maire, Michael and Belongie, Serge and Hays, James and Perona, Pietro and Ramanan, Deva and Doll{\'a}r, Piotr and Zitnick, C Lawrence},
  booktitle={European conference on computer vision},
  pages={740--755},
  year={2014},
  organization={Springer}
}

@article{chen2015microsoft,
  title={Microsoft coco captions: Data collection and evaluation server},
  author={Chen, Xinlei and Fang, Hao and Lin, Tsung-Yi and Vedantam, Ramakrishna and Gupta, Saurabh and Doll{\'a}r, Piotr and Zitnick, C Lawrence},
  journal={arXiv preprint arXiv:1504.00325},
  year={2015}
}

@article{young2014image,
  title={From image descriptions to visual denotations: New similarity metrics for semantic inference over event descriptions},
  author={Young, Peter and Lai, Alice and Hodosh, Micah and Hockenmaier, Julia},
  journal={Transactions of the association for computational linguistics},
  volume={2},
  pages={67--78},
  year={2014},
  publisher={MIT Press One Rogers Street, Cambridge, MA 02142-1209, USA journals-info~…}
}

@misc{nips-2017-defense-against-adversarial-attack,
    author = {Alex K and Ben Hamner and Ian Goodfellow},
    title = {NIPS 2017: Defense Against Adversarial Attack},
    year = {2017},
    howpublished = {\url{https://kaggle.com/competitions/nips-2017-defense-against-adversarial-attack}},
    note = {Kaggle}
}

@article{liu2024autodan,
  title={Autodan-turbo: A lifelong agent for strategy self-exploration to jailbreak llms},
  author={Liu, Xiaogeng and Li, Peiran and Suh, Edward and Vorobeychik, Yevgeniy and Mao, Zhuoqing and Jha, Somesh and McDaniel, Patrick and Sun, Huan and Li, Bo and Xiao, Chaowei},
  journal={arXiv preprint arXiv:2410.05295},
  year={2024}
}

@article{peng2023study,
  title={A study of generative large language model for medical research and healthcare},
  author={Peng, Cheng and Yang, Xi and Chen, Aokun and Smith, Kaleb E and PourNejatian, Nima and Costa, Anthony B and Martin, Cheryl and Flores, Mona G and Zhang, Ying and Magoc, Tanja and others},
  journal={NPJ digital medicine},
  volume={6},
  number={1},
  pages={210},
  year={2023},
  publisher={Nature Publishing Group UK London}
}

@article{chen2025janus,
  title={Janus-pro: Unified multimodal understanding and generation with data and model scaling},
  author={Chen, Xiaokang and Wu, Zhiyu and Liu, Xingchao and Pan, Zizheng and Liu, Wen and Xie, Zhenda and Yu, Xingkai and Ruan, Chong},
  journal={arXiv preprint arXiv:2501.17811},
  year={2025}
}

@article{feng2026backdooragent,
  title={Backdooragent: A unified framework for backdoor attacks on llm-based agents},
  author={Feng, Yunhao and Li, Yige and Wu, Yutao and Tan, Yingshui and Guo, Yanming and Ding, Yifan and Zhai, Kun and Ma, Xingjun and Jiang, Yu-Gang},
  journal={arXiv preprint arXiv:2601.04566},
  year={2026}
}

@inproceedings{zhou2017scene,
  title={Scene parsing through ade20k dataset},
  author={Zhou, Bolei and Zhao, Hang and Puig, Xavier and Fidler, Sanja and Barriuso, Adela and Torralba, Antonio},
  booktitle={Proceedings of the IEEE conference on computer vision and pattern recognition},
  pages={633--641},
  year={2017}
}

@inproceedings{cordts2016cityscapes,
  title={The cityscapes dataset for semantic urban scene understanding},
  author={Cordts, Marius and Omran, Mohamed and Ramos, Sebastian and Rehfeld, Timo and Enzweiler, Markus and Benenson, Rodrigo and Franke, Uwe and Roth, Stefan and Schiele, Bernt},
  booktitle={Proceedings of the IEEE conference on computer vision and pattern recognition},
  pages={3213--3223},
  year={2016}
}

@article{ke2023segment,
  title={Segment anything in high quality},
  author={Ke, Lei and Ye, Mingqiao and Danelljan, Martin and Tai, Yu-Wing and Tang, Chi-Keung and Yu, Fisher and others},
  journal={Advances in Neural Information Processing Systems},
  volume={36},
  pages={29914--29934},
  year={2023}
}

@inproceedings{fang2024clip,
  title={Clip-guided generative networks for transferable targeted adversarial attacks},
  author={Fang, Hao and Kong, Jiawei and Chen, Bin and Dai, Tao and Wu, Hao and Xia, Shu-Tao},
  booktitle={European Conference on Computer Vision},
  pages={1--19},
  year={2024},
  organization={Springer}
}

@article{dong2023robust,
  title={How robust is google's bard to adversarial image attacks?},
  author={Dong, Yinpeng and Chen, Huanran and Chen, Jiawei and Fang, Zhengwei and Yang, Xiao and Zhang, Yichi and Tian, Yu and Su, Hang and Zhu, Jun},
  journal={arXiv preprint arXiv:2309.11751},
  year={2023}
}

@inproceedings{naseer2021generating,
  title={On generating transferable targeted perturbations},
  author={Naseer, Muzammal and Khan, Salman and Hayat, Munawar and Khan, Fahad Shahbaz and Porikli, Fatih},
  booktitle={Proceedings of the IEEE/CVF international conference on computer vision},
  pages={7708--7717},
  year={2021}
}

@inproceedings{mahmood2021robustness,
  title={On the robustness of vision transformers to adversarial examples},
  author={Mahmood, Kaleel and Mahmood, Rigel and Van Dijk, Marten},
  booktitle={Proceedings of the IEEE/CVF international conference on computer vision},
  pages={7838--7847},
  year={2021}
}

@inproceedings{wang2023towards,
  title={Towards transferable targeted adversarial examples},
  author={Wang, Zhibo and Yang, Hongshan and Feng, Yunhe and Sun, Peng and Guo, Hengchang and Zhang, Zhifei and Ren, Kui},
  booktitle={Proceedings of the IEEE/CVF conference on computer vision and pattern recognition},
  pages={20534--20543},
  year={2023}
}

@inproceedings{wu2024improving,
  title={Improving transferable targeted adversarial attacks with model self-enhancement},
  author={Wu, Han and Ou, Guanyan and Wu, Weibin and Zheng, Zibin},
  booktitle={Proceedings of the IEEE/CVF Conference on Computer Vision and Pattern Recognition},
  pages={24615--24624},
  year={2024}
}

@inproceedings{wei2023enhancing,
  title={Enhancing the self-universality for transferable targeted attacks},
  author={Wei, Zhipeng and Chen, Jingjing and Wu, Zuxuan and Jiang, Yu-Gang},
  booktitle={Proceedings of the IEEE/CVF conference on computer vision and pattern recognition},
  pages={12281--12290},
  year={2023}
}

@article{shao2021adversarial,
  title={On the adversarial robustness of vision transformers},
  author={Shao, Rulin and Shi, Zhouxing and Yi, Jinfeng and Chen, Pin-Yu and Hsieh, Cho-Jui},
  journal={arXiv preprint arXiv:2103.15670},
  year={2021}
}

@inproceedings{wei2022towards,
  title={Towards transferable adversarial attacks on vision transformers},
  author={Wei, Zhipeng and Chen, Jingjing and Goldblum, Micah and Wu, Zuxuan and Goldstein, Tom and Jiang, Yu-Gang},
  booktitle={Proceedings of the AAAI conference on artificial intelligence},
  volume={36},
  number={3},
  pages={2668--2676},
  year={2022}
}

@article{zheng2024black,
  title={Black-box targeted adversarial attack on segment anything (sam)},
  author={Zheng, Sheng and Zhang, Chaoning and Hao, Xinhong},
  journal={IEEE Transactions on Multimedia},
  volume={27},
  pages={1901--1913},
  year={2024},
  publisher={IEEE}
}

@article{ravi2024sam,
  title={Sam 2: Segment anything in images and videos},
  author={Ravi, Nikhila and Gabeur, Valentin and Hu, Yuan-Ting and Hu, Ronghang and Ryali, Chaitanya and Ma, Tengyu and Khedr, Haitham and R{\"a}dle, Roman and Rolland, Chloe and Gustafson, Laura and others},
  journal={arXiv preprint arXiv:2408.00714},
  year={2024}
}

@inproceedings{vedantam2015cider,
  title={Cider: Consensus-based image description evaluation},
  author={Vedantam, Ramakrishna and Lawrence Zitnick, C and Parikh, Devi},
  booktitle={Proceedings of the IEEE conference on computer vision and pattern recognition},
  pages={4566--4575},
  year={2015}
}
}
\newpage
\appendix
\section{Impact Statements} 
Our DarkLLM imposes several positive broader impacts. \textbf{1)} DarkLLM presents an elegant and versatile framework for generating adversarial examples by leveraging the semantic reasoning of Large Language Models. This instruction-guided paradigm may inspire researchers to explore similar cross-modal approaches, utilizing the reasoning capabilities of LLMs to probe vulnerabilities in other domains, such as audio or video understanding. \textbf{2)} DarkLLM provides the first unified benchmark for evaluating targeted, untargeted, and multi-model attacks within a single system, paving the way for future research into language-driven safety assessments. \textbf{3)} Furthermore, by successfully crafting perturbations that simultaneously compromise both CLIP and SAM, DarkLLM validates the feasibility of cross-task adversarial transferability. This discovery suggests that functionally distinct foundation models share fundamental semantic vulnerabilities, encouraging researchers to explore even broader universal perturbations capable of spanning wider arrays of model families. \textbf{4)} Moreover, by establishing a robust pipeline for generating pixel-level perturbations via LLMs, DarkLLM opens new avenues for synergistic attacks. Researchers may explore integrating these visual perturbations with text-based jailbreak techniques, potentially unlocking more potent, multimodal adversarial strategies that exploit vulnerabilities across both visual and textual modalities simultaneously.
\section{Ethical Considerations}
We acknowledge the dual-use nature of DarkLLM, which, while designed for safety evaluation, could potentially be repurposed for malicious intent. To ensure responsible usage, we emphasize that DarkLLM is intended strictly for two legitimate purposes: (i) Red-Teaming: assisting developers in stress-testing the robustness of foundation models to identify and mitigate vulnerabilities; and (ii) Privacy Preservation: empowering individuals to shield their personal data from unauthorized analysis by ubiquitous vision systems. We strictly prohibit the use of this framework for any illegal activities, including disrupting lawful services or evading legitimate content moderation. By sharing our methodology, we aim to equip the research community with the insights necessary to build more resilient defenses against the emerging generation of language-driven adversarial threats, prioritizing the advancement of AI security over offensive proliferation.

\begin{table}[h]
    \centering
    \caption{A summary of our considered evaluation tasks, datasets, models, and performance metrics.}
    \resizebox{\columnwidth}{!}{
    \begin{tabular}{l|c|c|c|c}
    \toprule[1.5pt]
    \textbf{Evaluation Task} & \textbf{Attack Objective} & \textbf{Evaluation Model} & \textbf{Evaluation Dataset} & \textbf{Evaluation Metric}  \\
    \midrule[0.75pt]
    Zero-Shot Classification & Untargeted 
    & \makecell[c]{CLIP-RN101, CLIP-ViT-B-16\\CLIP-ViT-B-32, CLIP-ViT-L-14}  
    & \makecell[c]{CIFAR10, CIFAR100, Food\\GTSRB, ImageNet, Cars, STL} 
    & ASR  \\
    \midrule[0.75pt]
    Image-Text Retrieval & Untargeted 
    & \makecell[c]{CLIP-RN101, CLIP-ViT-B-16\\CLIP-ViT-B-32, CLIP-ViT-L-14}
    & MS-COCO 
    & ASR  \\
    \midrule[0.75pt]
    Image Captioning & Untargeted 
    & LLaVA-7B 
    & MS-COCO, Flickr30k 
    & ASR  \\
    \midrule[0.75pt]
    Image Captioning & Targeted 
    & \makecell[c]{Qwen3-VL-8B, GPT-4o\\GPT-4.1, GPT-5-mini, Gemini-2.0-Flash\\Gemini-3.0-Flash} 
    & NIPS 2017 Competition 
    & ASR  \\
    \midrule[0.75pt]
    Image Segmentation & Untargeted 
    & \makecell[c]{SAM-B, SAM-H\\HQ-SAM-H, SAM2-Hiera-Large}   
    & COCO, ADE, Cityscapes 
    & mIoU, ASR  \\
    \bottomrule[1.5pt]
    \end{tabular}
    }
    \label{tab:task_summary}
\end{table}

\section{Experiments}
\label{sec:supp-exps}
\subsection{Detailed Experimental Setting}
\noindent\textbf{More Implementation Details.}\quad 
We implement DarkLLM using the XTuner codebase~\cite{2023xtuner} on 8 NVIDIA H200 GPUs. The LLM controller is initialized from InternVL2.5-1B~\cite{chen2024expanding} and fine-tuned via LoRA~\cite{hu2022lora} ($r=128, \alpha=256$). Optimization is performed using AdamW with $\beta=(0.9, 0.999)$ and a weight decay of $0.05$. We utilize a compound learning rate scheduler: a linear warmup phase for the initial 5\% of iterations, followed by a cosine annealing decay for the remainder.

Our training follows a decoupled progressive strategy designed to mitigate the convergence imbalance between different attack objectives. As detailed in Table~\ref{tab:input_qa_appendix}, we assign unique control tokens (e.g., \texttt{[VLM\_TGT]}, \texttt{[SEG\_ADV]}) to route instructions to their dedicated noise generators, effectively isolating gradient updates for incompatible tasks.
\textit{Stage 1 (Targeted Instruction Tuning):} We first jointly train the LLM controller and the targeted generator using ImageNet~\cite{deng2009imagenet} (55 epochs) and CC3M~\cite{sharma2018conceptual} (10 epochs), with a per-GPU batch size of 128. To construct the training QA pairs, we use images from the COCO 2017 Caption training~\cite{lin2014microsoft} set as the target images, and leverage their corresponding ground-truth captions to populate the \{target\_caption\} placeholder in our instructional template (see Table~\ref{tab:input_qa_appendix}). This stage focuses exclusively on mastering the semantic parsing of complex targeted instructions.
\textit{Stage 2 (Unified Attack Training):} We then initialize the system with Stage-1 weights. In this phase, we freeze the targeted generator to preserve its learned capabilities and introduce the untargeted and multi-objective generators, which are optimized alongside the LLM controller. This stage involves joint training on all attack types for 8 epochs with a batch size of 64, enabling the LLM to learn unified control across diverse adversarial tasks.

Leveraging our modular design, we flexibly assign perturbation budgets ($\epsilon$) tailored to each task's standard benchmarks. Specifically, we set $\epsilon=16/255$ for targeted MLLM attacks following the protocol of AnyAttack~\cite{zhang2025anyattack}. For SAM attacks, we also use $\epsilon=10/255$, and all baselines are evaluated under identical settings to ensure fair comparison. For untargeted VLM attacks, we use $\epsilon=12/255$ following X-Transfer~\cite{huang2025x}. Finally, for multi-model attacks, we adopt $\epsilon=12/255$ to ensure optimal attack performance. The unified training objective balances the active surrogate losses:
\begin{equation}
\mathcal{L}_{\text{total}}
= \lambda_{\text{t}}\mathcal{L}_{\text{txt}}
+ \lambda_{\text{s}}\mathcal{L}_{\text{seg}}
+ \lambda_{\text{ta}}\mathcal{L}_{\text{target}}
+ \lambda_{\text{un}}\mathcal{L}_{\text{untarget}},
\label{eq:total_loss}
\end{equation}
\noindent where $\lambda_{\text{t}}=1.0$, $\lambda_{\text{s}}=10.0$, $\lambda_{\text{ta}}=10.0$, and $\lambda_{\text{un}}=8.0$, respectively, to balance different attack objectives.

\begin{table}[t]
\centering
\caption{Example of our training QA template. \{target\_caption\} is replaced by the actual caption during training.  }
\begin{minipage}{0.98\textwidth}\vspace{0mm}    \centering
\begin{tcolorbox} 
    \raggedright
    \small
     \hspace{-6mm}

    $\texttt{Human}:$ Make the vision-language model identify this image as [\{target\_caption\}].  \PredSty{\texttt{<STOP>}} \\
    $\texttt{DarkLLM}$: Sure, the generated perturbation can be represented as \texttt{[VLM\_TGT]}.
    \PredSty{\texttt{<STOP>}} \\
        $\texttt{Human}:$ Make the vision-language model fail to recognize this image.  \PredSty{\texttt{<STOP>}} \\
    $\texttt{DarkLLM}$: Sure, the generated perturbation can be represented as \texttt{[VLM\_ADV]}.
    \PredSty{\texttt{<STOP>}} \\
        $\texttt{Human}:$ Make the segmentation model fail to segment this image.  \PredSty{\texttt{<STOP>}} \\
    $\texttt{DarkLLM}$: Sure, the generated perturbation can be represented as \texttt{[SEG\_ADV]}.
    \PredSty{\texttt{<STOP>}} \\
        $\texttt{Human}:$ Make the segmentation model fail to segment this image and the vision-language model fail to recognize this image simultaneously.  \PredSty{\texttt{<STOP>}} \\
    $\texttt{DarkLLM}$: Sure, the generated perturbation can be represented as \texttt{[MULTI\_ADV]}.
    \PredSty{\texttt{<STOP>}} \\ 

\end{tcolorbox}
    
\vspace{-2mm}
    \label{tab:input_qa_appendix}
\vspace{-5mm}
\end{minipage}
\end{table}

\begin{table}[!t]
\centering
\caption{Details of baseline attacks on vision–language models.}
\begin{adjustbox}{width=0.95\linewidth}
\begin{tabular}{@{}c|c|c@{}}
\toprule[1.5pt]
Method & Variant & Note \\ \midrule[0.75pt]
\multirow{2}{*}{GD-UAP} & Seg & Generated with segmentation model (ResNet152 backbone) as the surrogate. \\
 & CLS & Generated with ResNet152 classifier as the surrogate. \\ \midrule[0.75pt]
\multirow{2}{*}{TRM-UAP} & GoogleNet & Generated with GoogleNet classifier trained on ImageNet. \\
 & RN152 & Generated with RN152 classifier trained on ImageNet. \\ \midrule[0.75pt]
\multirow{2}{*}{Meta-UAP} & Ensemble & Generated with an ensemble of DenseNet121, VGG16, and ResNet50 classifiers trained on ImageNet. \\
 & Ensemble-Meta & Same as above, with meta-learning strategy. \\ \midrule[0.75pt]
\multirow{2}{*}{C-GPC} & RN101-COCO & Generated with CLIP ResNet101 released by OpenAI on the MSCOCO dataset. \\
 & ViT-B/16-COCO & Generated with CLIP ViT-B/16 released by OpenAI on the MSCOCO dataset. \\ \midrule[0.75pt]
\multirow{2}{*}{ETU} & RN101-Flicker & Generated with CLIP ResNet101 released by OpenAI on the Flicker dataset. \\
 & ViT-B/16-Flicker & Generated with CLIP ViT-B/16 released by OpenAI on the Flicker dataset. \\ \midrule[0.75pt]
\multirow{2}{*}{X-Transfer} & Vanilla & Generated with multiple CLIP encoders without surrogate scaling. \\
 & Base & Generated with 16 CLIP encoders under the surrogate scaling strategy. \\
  \bottomrule[1.5pt]
\end{tabular}
\end{adjustbox}
\label{tab:clip_baselines}
\end{table}

\noindent\textbf{Evaluation Baselines.} We benchmark DarkLLM against state-of-the-art baselines across three distinct settings. \textbf{(1)} For untargeted attacks against VLMs, we compare our approach with state-of-the-art UAP methods specifically tailored for CLIP encoders, including X-Transfer~\cite{huang2025x}, C-PGC~\cite{fang2025one}, and ETU~\cite{zhang2024universal}. We also include foundational UAP methods designed for image classifiers, such as GD-UAP~\cite{mopuri2018generalizable}, TRM-UAP~\cite{liu2023trm}, and Meta-UAP~\cite{weng2024learning}. The details of each baseline are summarized in Table~\ref{tab:clip_baselines}. \textbf{(2)} For targeted attacks against MLLMs, we conduct a systematic comparison against two leading categories of methods: the sample-wise approaches AttackVLM~\cite{zhao2023evaluating} and M-Attack~\cite{li2025frustratingly}, and the universal attack method AnyAttack~\cite{zhang2025anyattack}. The details of each baseline are summarized in Table~\ref{tab:mllm_baselines}. \textbf{(3)} For untargeted attacks against SAM, our comparison includes leading sample-wise methods Attack-SAM~\cite{zhang2023attack}, PATA~\cite{zheng2024black}, and UAD~\cite{lu2024unsegment}, as well as the universal methods DarkSAM~\cite{zhou2024darksam} and UAP-SAM2~\cite{zhou2025vanish}. The details of each baseline are summarized in Table~\ref{tab:sam_baselines}.

\begin{table}[h]
\centering
\caption{Details of baseline attacks on multimodal large language models.}
\begin{adjustbox}{width=0.95\linewidth}
\begin{tabular}{@{}c|c|c@{}}
\toprule[1.5pt]
Method & Variant & Note \\ \midrule[0.75pt]
\multirow{3}{*}{AttackVLM} & B/16 & Generated with CLIP ViT-B/16 released by OpenAI under a perturbation budget of $\epsilon = 16/255$. \\
 & L/14 & Generated with CLIP-ViT-L-14-laion2B-s32B-b82K under a perturbation budget of $\epsilon = 16/255$. \\ 
  & Ensemble & Generated by jointly optimizing against CLIP ViT-B/16 and ViT-L/14 under $\epsilon = 16/255$. \\ \midrule[0.75pt]
\multirow{3}{*}{M-Attack} & B/16 & Generated with CLIP ViT-B/16 released by OpenAI under a perturbation budget of $\epsilon = 16/255$. \\
 & L/14 & Generated with CLIP-ViT-L-14-laion2B-s32B-b82K under a perturbation budget of $\epsilon = 16/255$. \\ 
  & Ensemble & Generated by jointly optimizing against CLIP ViT-B/16 and ViT-L/14 under $\epsilon = 16/255$. \\ \midrule[0.75pt]
\multirow{8}{*}{AnyAttack} & coco-bi & \makecell[c]{Generated by jointly optimizing against CLIP ViT-B/32 (OpenAI), \\ViT-B/16 (ImageNet), and ViT-L/14 EVA on MSCOCO under $\epsilon = 16/255$ using a contrastive loss.}  \\
 & coco-cos & \makecell[c]{Generated by jointly optimizing against CLIP ViT-B/32 (OpenAI), \\ViT-B/16 (ImageNet), and ViT-L/14 EVA on MSCOCO under $\epsilon = 16/255$ using a cosine similarity loss.} \\
  & flickr-bi & \makecell[c]{Generated by jointly optimizing against CLIP ViT-B/32 (OpenAI), \\ViT-B/16 (ImageNet), and ViT-L/14 EVA on Flickr dataset under $\epsilon = 16/255$ using a contrastive loss.} \\ 
   & flickr-cos & \makecell[c]{Generated by jointly optimizing against CLIP ViT-B/32 (OpenAI), \\ViT-B/16 (ImageNet), and ViT-L/14 EVA on Flickr dataset under $\epsilon = 16/255$ using a cosine similarity loss.} \\
  \bottomrule[1.5pt]
\end{tabular}
\end{adjustbox}
\label{tab:mllm_baselines}
\end{table}

\begin{table}[h]
\centering
\caption{Details of baseline attacks on segment anything models.}
\begin{adjustbox}{width=0.95\linewidth}
\begin{tabular}{@{}c|c|c@{}}
\toprule[1.5pt]
Method & Variant & Note \\ \midrule[0.75pt]
Attack-SAM   
    & --- & Generated with SAM-Base released by Meta under a perturbation budget of $\epsilon = 10/255$. \\ \midrule[0.75pt]
UAD   
    & --- & Generated with SAM-Base released by Meta under a perturbation budget of $\epsilon = 10/255$. \\ \midrule[0.75pt]
    PATA   
    & --- & Generated with SAM-Base released by Meta under a perturbation budget of $\epsilon = 10/255$. \\ \midrule[0.75pt]
        PATA++
    & --- & Generated with SAM-Base released by Meta under a perturbation budget of $\epsilon = 10/255$. \\ \midrule[0.75pt]
\multirow{2}{*}{DarkSAM}   
    & DarkSAM-pt & Generated with SAM-Base on the SA-1B dataset under a perturbation budget of $\epsilon = 10/255$ using point prompts. \\
    & DarkSAM-bx & Generated with SAM-Base on the SA-1B dataset under a perturbation budget of $\epsilon = 10/255$ using box prompts. \\ \midrule[0.75pt]
UAP-SAM2   
    & --- & Generated with SAM2-Small on the DAVIS dataset under a perturbation budget of $\epsilon = 16/255$ using point prompts. \\ 
  \bottomrule[1.5pt]
\end{tabular}
\end{adjustbox}
\label{tab:sam_baselines}
\end{table}

\noindent\textbf{Evaluation Datasets.} 
We provide a comprehensive summary of our experimental setup, including evaluation tasks, models, datasets, and metrics, in Table~\ref{tab:task_summary}.
Our evaluation spans a wide range of standard benchmarks tailored to each adversarial task. \textbf{(1)} For untargeted attacks on VLMs, our experiments cover three task families: \textit{(i) Zero-shot Classification}, for which we use seven datasets (CIFAR-10, CIFAR-100~\cite{krizhevsky2009learning}, Food-101~\cite{bossard2014food}, GTSRB~\cite{stallkamp2012man}, ImageNet~\cite{deng2009imagenet}, Stanford Cars~\cite{krause2013collecting}, and STL-10~\cite{coates2011analysis}); \textit{(ii) Image-Text Retrieval}, using the COCO-Karpathy split~\cite{karpathy2015deep}; and \textit{(iii) Image Captioning}, evaluated on MS-COCO Caption~\cite{chen2015microsoft} and Flickr30k~\cite{young2014image}. \textbf{(2)} For targeted attacks on MLLMs, we follow prior work~\cite{dong2023robust,li2025frustratingly,jia2025adversarial} by using a subset of 100 images from the NIPS 2017 Adversarial Attacks and Defenses Competition dataset\footnote{\url{https://nips.cc/Conferences/2017/CompetitionTrack}}~\cite{nips-2017-defense-against-adversarial-attack}. \textbf{(3)} For untargeted attacks on SAM, we employ three widely-used segmentation benchmarks in the domain: ADE20K-val~\cite{zhou2017scene}, COCO2017-val~\cite{lin2014microsoft}, and Cityscapes-val~\cite{cordts2016cityscapes}.

\newcolumntype{P}[1]{>{\raggedright\arraybackslash}p{#1}}

\begin{wraptable}{r}{0.65\textwidth} 
\vspace{-8pt}
\centering
\caption{Encoders used in our zero-shot classification and image--text retrieval evaluations. For each encoder, we report its architecture, pre-training dataset, and the corresponding OpenCLIP identifier (i.e., the \texttt{model\_name} and \texttt{pretrained} arguments in OpenCLIP's \texttt{create\_model\_and\_transforms}).}
\label{tab:clip_evaluators}
\vspace{-6pt}
\begin{tabular*}{\linewidth}{@{\extracolsep{\fill}}c c c P{0.42\linewidth}@{}}
\toprule
 & Architecture & Pre-training Dataset & OpenCLIP Identifier \\ \midrule
1 & RN101      & WebImageText & (RN101, openai) \\
2 & ViT-B/16  & WebImageText & (ViT-B-16, openai) \\
3 & ViT-B/32  & WebImageText & (ViT-B-32, openai) \\
4 & ViT-L/14  & WebImageText & (ViT-L-14, openai) \\
\bottomrule
\end{tabular*}

\vspace{-10pt}
\end{wraptable}

\noindent\textbf{Evaluation Models.} 
We benchmark DarkLLM's transferability against a diverse set of foundation models. \textbf{(1)} For untargeted VLM attacks, performance on classification and retrieval is averaged across four black-box CLIP models as shown in Table~\ref{tab:clip_evaluators}, while image captioning is evaluated on LLaVA-1.5-7B~\cite{liu2023visual}. \textbf{(2)} For targeted MLLM attacks, our evaluation targets six frontier commercial models: Qwen3-VL, GPT-4o, GPT-4.1, GPT-5-mini, Gemini-2.0-Flash, and Gemini-3.0-Flash. \textbf{(3)} For untargeted SAM attacks, we assess cross-model transferability across the SAM family, including SAM-Base, SAM-Huge~\cite{kirillov2023segment}, HQ-SAM-Huge~\cite{ke2023segment}, and SAM2-Hiera-Large~\cite{ravi2024sam}.

\noindent\textbf{Evaluation Metrics.} 
We adopt task-specific metrics to comprehensively evaluate performance. \textbf{(1)} For untargeted VLM attacks, we report a normalized Attack Success Rate (ASR) defined as $(s_{\text{clean}} - s_{\text{adv}}) / s_{\text{clean}}$, where $s$ is the primary metric for each task (i.e., accuracy for classification, CIDEr~\cite{vedantam2015cider} for captioning). $s_{\text{clean}}$ and $s_{\text{adv}}$ are the scores on clean and adversarial images, respectively. \textbf{(2)} For targeted MLLM attacks, we follow the LLM-as-a-judge protocol following~\cite{li2025frustratingly,jia2025adversarial}. Specifically, we prompt the target MLLM itself to generate captions for both adversarial and target images, then assess their similarity via GPTScore. An attack is deemed successful if the similarity exceeds 0.3. We report both ASR and the average similarity score (AvgSim). Detailed evaluation prompts are provided in Table~\ref{tab:evaluation_prompt}. \textbf{(3)} For untargeted SAM attacks, we report both ASR and the standard mean Intersection over Union (mIoU) to measure segmentation quality degradation.

\begin{table}[!t]
\centering

\caption{Details of our evaluation prompt template. During evaluation, \textcolor{red}{text1} and \textcolor{red}{text2} are replaced with the clean prediction and the adversarial prediction, respectively.  }
\begin{minipage}{0.98\textwidth}\vspace{0mm}    \centering
\begin{tcolorbox} 
    \raggedright
    \small

    Rate the semantic similarity between the following two texts on a scale from 0 to 1.
        
                 **Criteria for similarity measurement:**
                    \\1. **Main Subject Consistency:** If both descriptions refer to the same key subject or object (e.g., a person, food, an event), they should receive a higher similarity score.
                    \\2. **Relevant Description**: If the descriptions are related to the same context or topic, they should also contribute to a higher similarity score.
                    \\3. **Ignore Fine-Grained Details:** Do not penalize differences in **phrasing, sentence structure, or minor variations in detail**. Focus on **whether both descriptions fundamentally describe the same thing.**
                    \\4. **Partial Matches:** If one description contains extra information but does not contradict the other, they should still have a high similarity score.
                   \\ 5. **Similarity Score Range:** 
                        \\\quad\quad\quad\quad- **1.0**: Nearly identical in meaning.
                       \\\quad\quad\quad\quad-- **0.8-0.9**: Same subject, with highly related descriptions.
                        \\\quad\quad\quad\quad-- **0.7-0.8**: Same subject, core meaning aligned, even if some details differ.
                        \\\quad\quad\quad\quad-- **0.5-0.7**: Same subject but different perspectives or missing details.
                        \\\quad\quad\quad\quad-- **0.3-0.5**: Related but not highly similar (same general theme but different descriptions).
                        \\\quad\quad\quad\quad-- **0.0-0.2**: Completely different subjects or unrelated meanings. \\
                    
                   Text 1: \{\textcolor{red}{text1}\}
                    \\ Text 2: \{\textcolor{red}{text2}\} \\

                 Output only a single number between 0 and 1. Do not include any explanation or additional text.

\end{tcolorbox}
    
    \label{tab:evaluation_prompt}
\end{minipage}
\end{table}

\noindent\textbf{Surrogate Models.} 
As the first framework designed to unify heterogeneous attack types, the optimization of DarkLLM leverages a diverse suite of surrogate models, which are strictly disjoint from the target models used for evaluation to ensure a fair black-box setting. \textbf{(1)} Specifically, for untargeted attacks on VLMs, we employ a surrogate scaling strategy utilizing an ensemble of 16 distinct CLIP variants to enhance perturbation transferability following~\cite{huang2025x}, with full details provided in Table~\ref{tab:base_search_space}. \textbf{(2)} For targeted attacks on MLLMs, our optimization relies on two powerful, publicly available CLIP models: the OpenAI ViT-B/16 and CLIP-ViT-L-14-laion2B-s32B-b82K. \textbf{(3)} Finally, for untargeted attacks against the SAM family, we use SAM-Base and the lightweight MobileSAM as our primary surrogates.

\definecolor{lightgray}{rgb}{0.90,0.90,0.90}   

\begin{table*}[htbp]
\setlength{\abovecaptionskip}{4pt}
  \centering
    \caption{The mIoU (\%) results for promptable image segmentation task across different SAM models.}
      \resizebox{0.8\textwidth}{!}{
\begin{tabular}{
  c|c|c c c|c c c|c c c
}
    \toprule[1.5pt]
    \multicolumn{2}{c|}{\textbf{Setting}} & \multicolumn{3}{c|}{\textbf{SAM-Base}} & \multicolumn{3}{c|}{\textbf{SAM-Large}}       & \multicolumn{3}{c}{\textbf{SAM-Huge}}     \\
 \cmidrule(lr){3-5} \cmidrule(lr){6-8} \cmidrule(lr){9-11}  \textbf{Prompt}  & \textbf{Method} & \texttt{\textbf{ADE}} & \texttt{\textbf{COCO}}  & \texttt{\textbf{CITY}}  & \texttt{\textbf{ADE}} & \texttt{\textbf{COCO}}  & \texttt{\textbf{CITY}}  & \texttt{\textbf{ADE}} & \texttt{\textbf{COCO}} & \texttt{\textbf{CITY}}  \\
    \midrule[0.75pt]
    \multirow{3}[2]{*}{Point} & Clean & 50.3 & 44.9 & 38.0 & 53.3 & 48.1 & 40.0 & 53.6 & 48.5 & 39.7    \\
          & UAD~\cite{lu2024unsegment} & 42.7  & 39.4  & 8.6 & 46.9 & 44.2 & 13.1 & 48.0  & 44.6  & 16.0   \\
          
          & \cellcolor{blue!5} DarkLLM (Ours) 
          & \cellcolor{blue!5}\textbf{2.9}     
          & \cellcolor{blue!5}\textbf{3.5}   
          & \cellcolor{blue!5}\textbf{0.4} 
          & \cellcolor{blue!5}\textbf{18.8}    
          & \cellcolor{blue!5}\textbf{22.0}     
          & \cellcolor{blue!5}\textbf{18.6}    
          & \cellcolor{blue!5}\textbf{21.5}   
          & \cellcolor{blue!5}\textbf{24.3}   
          & \cellcolor{blue!5}\textbf{19.2} 
          \\
    \midrule[0.75pt]
    \multirow{3}[2]{*}{Box } & Clean & 72.8 & 71.4 & 61.2 & 73.9 & 73.0 & 61.6 & 74.5 & 73.2 & 61.5     \\
          & UAD~\cite{lu2024unsegment} & 68.3  & 68.0  & 39.7 & 70.3 & 70.5 & 45.5 & 71.3 & 71.1 & 47.9     \\
          
          & \cellcolor{blue!5} DarkLLM (Ours) 
          & \cellcolor{blue!5}\textbf{27.4}      
          & \cellcolor{blue!5}\textbf{30.5}   
          & \cellcolor{blue!5}\textbf{30.2}     
          & \cellcolor{blue!5}\textbf{48.4}    
          & \cellcolor{blue!5}\textbf{55.5}    
          & \cellcolor{blue!5}\textbf{44.9}   
          & \cellcolor{blue!5}\textbf{46.9}   
          & \cellcolor{blue!5}\textbf{55.1}     
          & \cellcolor{blue!5}\textbf{43.4} \\
    \bottomrule[1.5pt]
    \end{tabular}%
    }
  \label{tab:appendix_sam}%
    \vspace{-0.2cm}
\end{table*}

\begin{table*}[htbp]
\setlength{\abovecaptionskip}{4pt}
  \centering
    \caption{The mIoU (\%) results for promptable image segmentation task across different HQ-SAM models.}
      \resizebox{\textwidth}{!}{
\begin{tabular}{
  c|c|c c c|c c c|c c c|c c c
}
    \toprule[1.5pt]
    \multicolumn{2}{c|}{\textbf{Setting}} & \multicolumn{3}{c|}{\textbf{HQ-SAM-Tiny}} & \multicolumn{3}{c|}{\textbf{HQ-SAM-Base}}       & \multicolumn{3}{c|}{\textbf{HQ-SAM-Large}}    & \multicolumn{3}{c}{\textbf{HQ-SAM-Huge}} \\
 \cmidrule(lr){3-5} \cmidrule(lr){6-8} \cmidrule(lr){9-11} \cmidrule(lr){12-14}  \textbf{Prompt}  & \textbf{Method} & \texttt{\textbf{ADE}} & \texttt{\textbf{COCO}}  & \texttt{\textbf{CITY}}  & \texttt{\textbf{ADE}} & \texttt{\textbf{COCO}}  & \texttt{\textbf{CITY}}  & \texttt{\textbf{ADE}} & \texttt{\textbf{COCO}} & \texttt{\textbf{CITY}}  & \texttt{\textbf{ADE}} & \texttt{\textbf{COCO}}  & \texttt{\textbf{CITY}} \\
    \midrule[0.75pt]
    \multirow{3}[2]{*}{Point} & Clean & 47.8 & 42.9 & 30.4 & 46.0 & 39.7 & 29.8 & 49.9 & 44.3 & 32.0 & 52.4 & 47.3 & 32.0   \\
          & UAD~\cite{lu2024unsegment} & 42.0 & 67.2 & 15.1 & 38.5 & 34.9 & 6.1 & 43.4 & 39.8 & 12.9 & 46.5  & 43.6  & 14.4  \\
          
          & \cellcolor{blue!5} DarkLLM (Ours) 
          & \cellcolor{blue!5}\textbf{15.0}     
          & \cellcolor{blue!5}\textbf{15.5}   
          & \cellcolor{blue!5}\textbf{16.4}     
          & \cellcolor{blue!5}\textbf{4.5}   
          & \cellcolor{blue!5}\textbf{5.2}   
          & \cellcolor{blue!5}\textbf{2.1} 
          & \cellcolor{blue!5}\textbf{17.0}
          & \cellcolor{blue!5}\textbf{19.3}
          & \cellcolor{blue!5}\textbf{14.6}
          & \cellcolor{blue!5}\textbf{18.9}    
          & \cellcolor{blue!5}\textbf{21.9}     
          & \cellcolor{blue!5}\textbf{14.2}
          \\
    \midrule[0.75pt]
    \multirow{3}[2]{*}{Box } & Clean & 71.4 & 69.8 & 57.6 & 66.3 & 60.6 & 46.6 & 70.1 & 66.8 & 50.2 & 69.4 & 65.2 & 50.2    \\
  
          & UAD~\cite{lu2024unsegment} & 68.3 & 67.2 & 50.2 & 61.1 & 58.4 & 18.1 & 66.1 & 64.4 & 37.9 & 65.8   & 63.7  & 33.5  \\
          
          & \cellcolor{blue!5} DarkLLM (Ours) 
          & \cellcolor{blue!5}\textbf{34.3}      
          & \cellcolor{blue!5}\textbf{37.7}   
          & \cellcolor{blue!5}\textbf{38.1}    
          & \cellcolor{blue!5}\textbf{32.0}   
          & \cellcolor{blue!5}\textbf{37.5}     
          & \cellcolor{blue!5}\textbf{28.3} 
          & \cellcolor{blue!5}\textbf{45.4}
          & \cellcolor{blue!5}\textbf{49.1} 
          & \cellcolor{blue!5}\textbf{35.8}   
          & \cellcolor{blue!5}\textbf{40.9}    
          & \cellcolor{blue!5}\textbf{47.1}    
          & \cellcolor{blue!5}\textbf{32.3} \\
    \bottomrule[1.5pt]
    \end{tabular}%
    }
  \label{tab:appendix_sam_hq}%
    \vspace{-0.2cm}
\end{table*}

\begin{table*}[t]
\setlength{\abovecaptionskip}{4pt}
  \centering
    \caption{The mIoU (\%) results for promptable image segmentation task across different SAM2 models.}
      \resizebox{\textwidth}{!}{
\begin{tabular}{
  c|c|c c c|c c c|c c c|c c c
}
    \toprule[1.5pt]
    \multicolumn{2}{c|}{\textbf{Setting}} & \multicolumn{3}{c|}{\textbf{SAM2-Hiera-Tiny}} & \multicolumn{3}{c|}{\textbf{SAM2-Hiera-Small}}       & \multicolumn{3}{c|}{\textbf{SAM2-Hiera-Base}}    & \multicolumn{3}{c}{\textbf{SAM2-Hiera-Large}} \\
 \cmidrule(lr){3-5} \cmidrule(lr){6-8} \cmidrule(lr){9-11} \cmidrule(lr){12-14}  \textbf{Prompt}  & \textbf{Method} & \texttt{\textbf{ADE}} & \texttt{\textbf{COCO}}  & \texttt{\textbf{CITY}}  & \texttt{\textbf{ADE}} & \texttt{\textbf{COCO}}  & \texttt{\textbf{CITY}}  & \texttt{\textbf{ADE}} & \texttt{\textbf{COCO}} & \texttt{\textbf{CITY}}  & \texttt{\textbf{ADE}} & \texttt{\textbf{COCO}}  & \texttt{\textbf{CITY}} \\
    \midrule[0.75pt]
    \multirow{3}[2]{*}{Point} & Clean & 52.8 & 49.6 & 37.3 & 53.5 & 49.5 & 23.1 & 55.4 & 51.0 & 40.6 & 57.2 & 55.2 & 41.7  \\
          & UAD~\cite{lu2024unsegment} & 48.1 & 48.2 & 25.0 & 48.9 & 48.3 & 14.2 & 50.5 & 50.0 & 23.2 & 52.6  & 52.7  & 26.0 \\
          
          & \cellcolor{blue!5} DarkLLM (Ours) 
          & \cellcolor{blue!5}\textbf{19.7}     
          & \cellcolor{blue!5}\textbf{21.4}   
          & \cellcolor{blue!5}\textbf{16.4}     
          & \cellcolor{blue!5}\textbf{18.9}   
          & \cellcolor{blue!5}\textbf{21.1}   
          & \cellcolor{blue!5}\textbf{16.7} 
          & \cellcolor{blue!5}\textbf{16.4}    
          & \cellcolor{blue!5}\textbf{19.8}     
          & \cellcolor{blue!5}\textbf{15.3}
          & \cellcolor{blue!5}\textbf{20.2}
          & \cellcolor{blue!5}\textbf{23.4}
          & \cellcolor{blue!5}\textbf{15.1}
          \\
    \midrule[0.75pt]
    \multirow{3}[2]{*}{Box } & Clean & 73.8 & 72.9 & 61.5 & 74.0 & 73.1 & 62.1 & 74.6 & 74.0 & 63.2 & 74.9 & 73.9 & 62.7   \\

          & UAD~\cite{lu2024unsegment} & 71.8 & 71.4 & 53.9 & 72.1 & 71.6 & 53.2 & 72.5 & 72.5 & 52.5 & 73.0  & 72.7  & 53.4   \\
          
          & \cellcolor{blue!5} DarkLLM (Ours) 
          & \cellcolor{blue!5}\textbf{55.9}      
          & \cellcolor{blue!5}\textbf{58.4}   
          & \cellcolor{blue!5}\textbf{45.9}    
          & \cellcolor{blue!5}\textbf{51.0}   
          & \cellcolor{blue!5}\textbf{55.3}     
          & \cellcolor{blue!5}\textbf{43.0}    
          & \cellcolor{blue!5}\textbf{50.1}    
          & \cellcolor{blue!5}\textbf{55.4}    
          & \cellcolor{blue!5}\textbf{43.2}
          & \cellcolor{blue!5}\textbf{49.5}
          & \cellcolor{blue!5}\textbf{55.7} 
          & \cellcolor{blue!5}\textbf{38.8} \\
    \bottomrule[1.5pt]
    \end{tabular}%
    }
  \label{tab:appendix_sam2}
    \vspace{-0.2cm}
\end{table*}

\subsection{Additional Experiments}
\noindent\textbf{Comprehensive Evaluations on the SAM.}\quad
We conduct a comprehensive evaluation to validate DarkLLM's effectiveness across the broader SAM family. Tables~\ref{tab:appendix_sam}, \ref{tab:appendix_sam_hq}, and \ref{tab:appendix_sam2} summarize the performance against various backbones of SAM~\cite{kirillov2023segment}, HQ-SAM~\cite{ke2023segment}, and SAM2~\cite{ravi2024sam}, respectively. Three key findings emerge from these results: (1) Across all tested models, datasets, and prompt types, DarkLLM consistently and substantially outperforms the previous state-of-the-art method, UAD. This demonstrates the superior adversarial efficacy and generalization of our approach. (2) While UAD shows strong performance on Cityscapes with point prompts, its effectiveness significantly degrades on other datasets and across all box prompt settings. In contrast, DarkLLM maintains robust performance in these scenarios, highlighting its superior transferability across both datasets and prompt modalities. (3) The results also confirm that box prompts are inherently more robust than point prompts, yielding higher clean mIoU scores. Despite this, DarkLLM effectively degrades segmentation performance even in these challenging settings, further validating its potency. We believe these comprehensive results offer valuable insights for designing both more advanced adversarial attacks and the corresponding defenses for large-scale segmentation models.

\noindent\textbf{Generalization Evaluation}\quad
DarkLLM can generate targeted visual perturbations based on user instructions. To demonstrate the generalization ability of our method to novel input instructions, we use Qwen3-VL 8B to generate new captions for the target images in the \texttt{NIPS 2017 Adversarial Attacks and Defenses Competition} dataset. As illustrated in Table~\ref{tab:caption_comparison}, these new captions provide more detailed and context-rich descriptions compared to the original annotations, thereby introducing diverse and previously unseen linguistic variations. These new captions are then incorporated into the instruction templates shown in Table~\ref{tab:input_qa_appendix} to prompt DarkLLM to generate targeted visual perturbations. As reported in Table~\ref{tab:attack_mllm_appendix}, the perturbations generated from these new captions achieve high ASR across different types of commercial models, further validating the robustness and generalization capability of our approach.

\begin{table}[h]
\caption{Comparison between original reference captions and captions generated by Qwen3-VL 8B.}
\centering
\begin{minipage}{0.98\textwidth}\vspace{0mm}
\centering
\begin{tcolorbox} 
    \raggedright
    \small
    \textbf{Original Captions:} \\
    (1) A dog chewing on a object held in a hand. \\
    (2) A group of young people getting ready to go ski. \\
    (3) A man that is laying down underneath a cat. \\
    \textbf{Qwen3-VL Generated Captions:} \\
    (1) A black and white dog rests its head on a person's lap while being gently petted with a red toy. \\
    (2) A group of skiers with large backpacks and poles stand on a snowy slope, adjusting gear before descent. \\
    (3) A man sleeps peacefully under white covers, while two cats rest nearby on pillows and the bedspread.
\end{tcolorbox}

\label{tab:caption_comparison}
\end{minipage}
\end{table}

\begin{table*}[t]
\centering
\vspace{-2mm}
\caption{Targeted ASR (\%) and AvgSim results for image captioning tasks across different commercial MLLMs.}
\resizebox{0.99\linewidth}{!}{ 
\begin{tabular}{l|c|cc|cc|cc|cc|cc}
\toprule[1.5pt]
& 
& \multicolumn{2}{c|}{\textbf{GPT-4o}} 
& \multicolumn{2}{c|}{\textbf{GPT-4.1}} 
& \multicolumn{2}{c|}{\textbf{GPT-5-mini}} 
& \multicolumn{2}{c|}{\textbf{Gemini-2.5-Flash}}
& \multicolumn{2}{c}{\textbf{Gemini-3-Flash}} \\
\cmidrule{3-4} \cmidrule{5-6} \cmidrule{7-8} \cmidrule{9-10} \cmidrule{11-12} 
\multirow{-2}{*}{\textbf{Method}} & \multirow{-2}{*}{\textbf{Base Model}} & \texttt{\textbf{ASR}} & \texttt{\textbf{AvgSim}} & \texttt{\textbf{ASR}} & \texttt{\textbf{AvgSim}} & \texttt{\textbf{ASR}} & \texttt{\textbf{AvgSim}} & \texttt{\textbf{ASR}} & \texttt{\textbf{AvgSim}} & \texttt{\textbf{ASR}} & \texttt{\textbf{AvgSim}} \\
\hline

\cellcolor{blue!10} DarkLLM (Ours) & 
\cellcolor{blue!10} InternVL-2B & 
\cellcolor{blue!10} {63.0} & 
\cellcolor{blue!10} {0.26} & 
\cellcolor{blue!10} {68.0} & 
\cellcolor{blue!10} {0.27} & 
\cellcolor{blue!10} {61.0} & 
\cellcolor{blue!10} {0.24} & 
\cellcolor{blue!10} {60.0} & 
\cellcolor{blue!10} {0.24} & 
\cellcolor{blue!10} {57.0} & 
\cellcolor{blue!10} {0.24} \\
\aboverulesepcolor{blue!10!}
\bottomrule[1.5pt]
\end{tabular}
} 
\vspace{-4mm}
\label{tab:attack_mllm_appendix}
\end{table*}

\begin{table}[t]
\centering
\caption{Surrogate encoders used for training, with architectures, pre-training datasets, and OpenCLIP identifiers.}
\label{tab:base_search_space}
\begin{adjustbox}{width=0.78\linewidth}
\begin{tabular}{@{}cccc@{}}
\toprule[1.5pt]
 & Architecture & Pre-training Dataset & OpenCLIP Identifier \\ \midrule[0.75pt]
1 & ConvNext Base & LAION400M & (convnext\_base, laion400m\_s13b\_b51k) \\
2 & ConvNext Base-W & LAION2B & (convnext\_base\_w, laion2b\_s13b\_b82k) \\
3 & ConvNext Base-W & LAION2B & (convnext\_base\_w, laion2b\_s13b\_b82k\_augreg) \\
4 & ConvNext Large-D & LAION2B & (convnext\_large\_d, laion2b\_s26b\_b102k\_augreg) \\

5 & ViT-B/16 & DFN2B & (ViT-B-16, dfn2b) \\
6 & ViT-B/16 & DataComp & (ViT-B-16, datacomp\_xl\_s13b\_b90k) \\
7 & ViT-B/16 & LAION2B & (ViT-B-16, laion2b\_s34b\_b88k) \\

8 & ViT-B/32 & LAION2B & (ViT-B-32, laion2b\_s34b\_b79k) \\
9 & ViT-B/32 & DataComp & (ViT-B-32, datacomp\_xl\_s13b\_b90k) \\
10 & ViT-B/32 & LAION400M & (ViT-B-32-quickgelu, laion400m\_e32) \\

11 & ViT-L/14 & LAION400M & (ViT-L-14, laion400m\_e32) \\
12 & ViT-L/14 & CommonPool & (ViT-L-14, commonpool\_xl\_s13b\_b90k) \\
13 & ViT-L/14 & MetaCLIP & (ViT-L-14-quickgelu, metaclip\_fullcc) \\
14 & ViT-L/14 & LAION2B & (ViT-L-14, laion2b\_s34b\_b88k) \\
15 & EVA02-L/14 & Merged2B & (EVA02-L-14, merged2b\_s4b\_b131k) \\
16 & ViT-L/14 & DataComp & (ViT-L-14, datacomp\_xl\_s13b\_b90k) \\ \bottomrule[1.5pt]
\end{tabular}
\end{adjustbox}
\end{table}

\begin{figure*}[h]
\begin{center}
\includegraphics[width=0.98\linewidth]{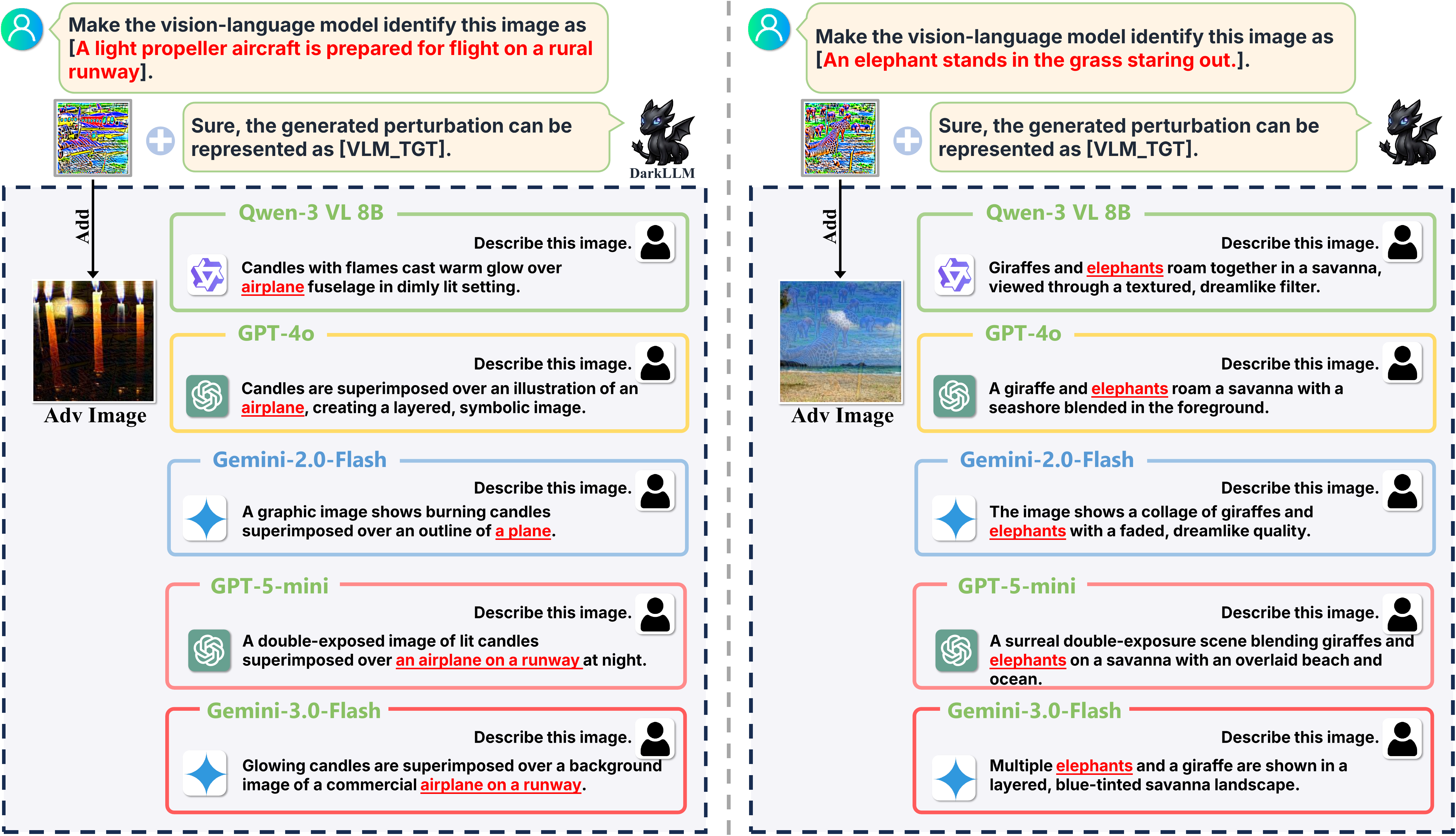}
\end{center}
    \caption{Visualization of DarkLLM for instruction-guided attacks on Commercial MLLMs.}
\label{fig:visualization_mllm}
\end{figure*}

\begin{figure*}[h]
\begin{center}
\includegraphics[width=0.98\linewidth]{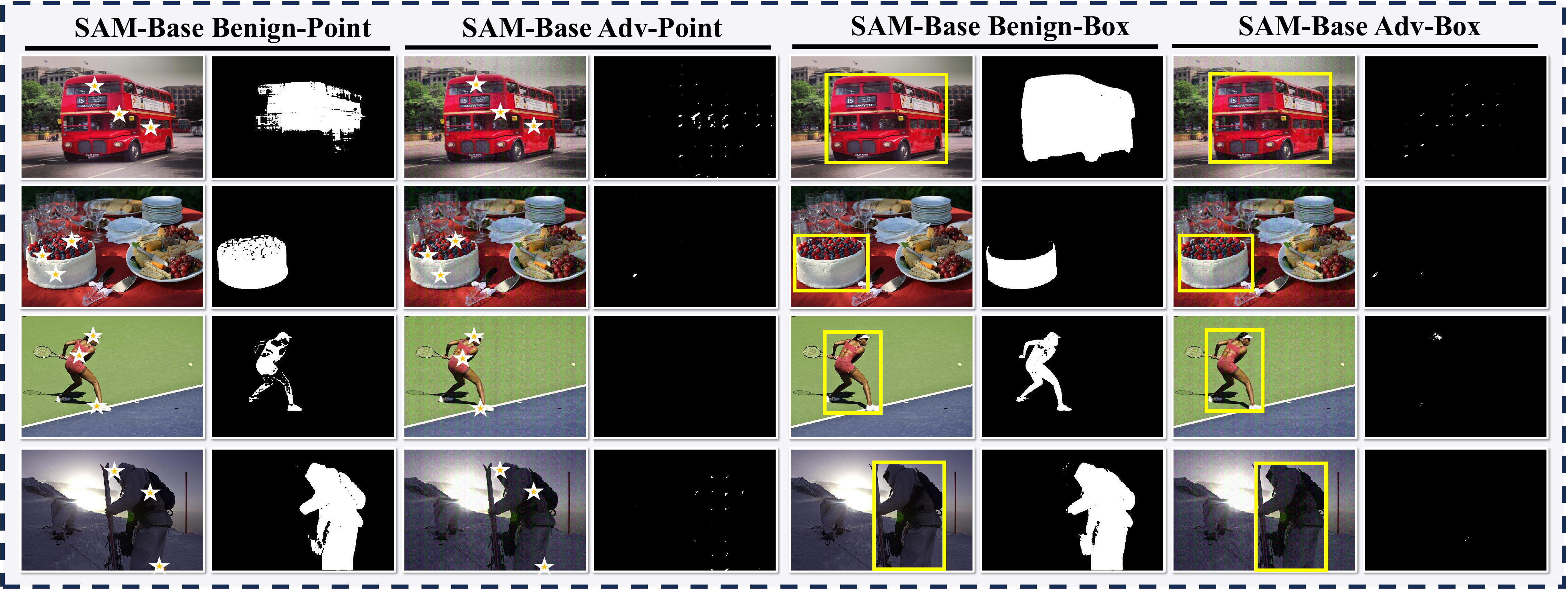}
\end{center}
    \caption{Visualization of DarkLLM for instruction-guided attacks on SAM-Base.}
\label{fig:visualization_sam_b}
\end{figure*}

\begin{figure*}[!t]
\begin{center}
\includegraphics[width=0.98\linewidth]{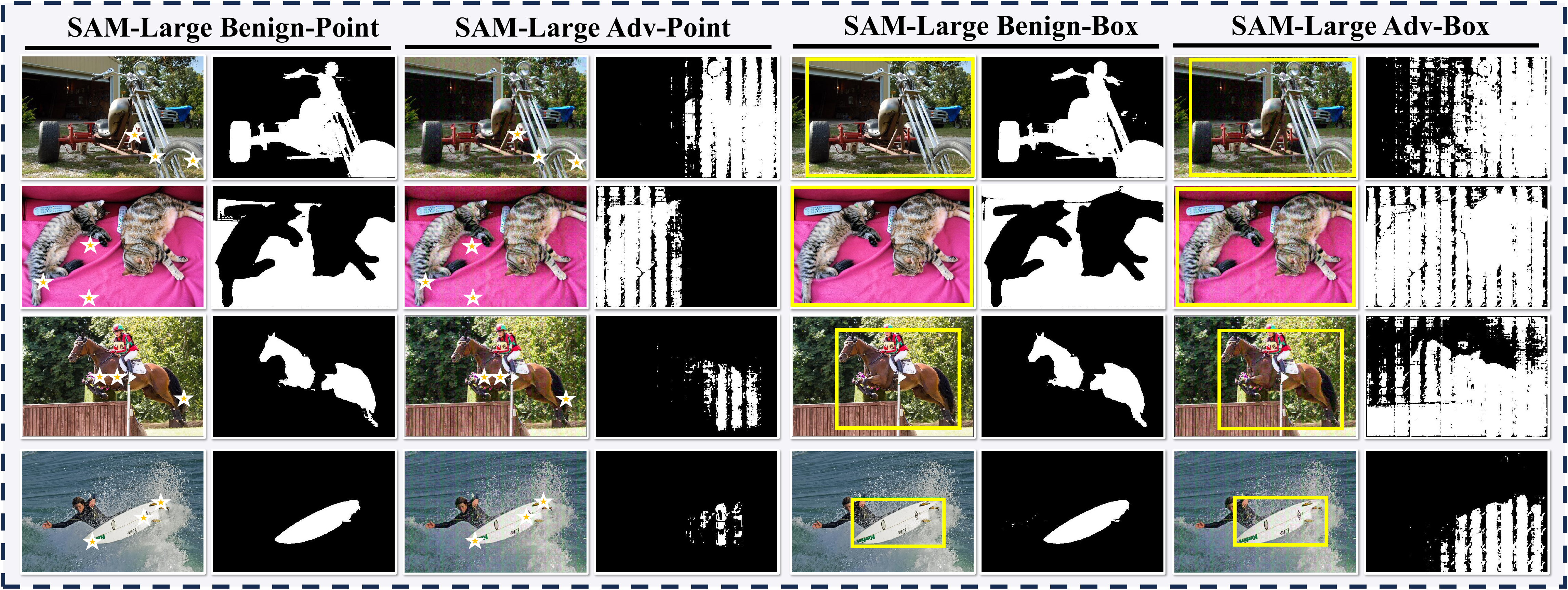}
\end{center}
    \caption{Visualization of DarkLLM for instruction-guided attacks on SAM-Large.}
\label{fig:visualization_sam_l}
\end{figure*}

\begin{figure*}[!t]
\begin{center}
\includegraphics[width=0.98\linewidth]{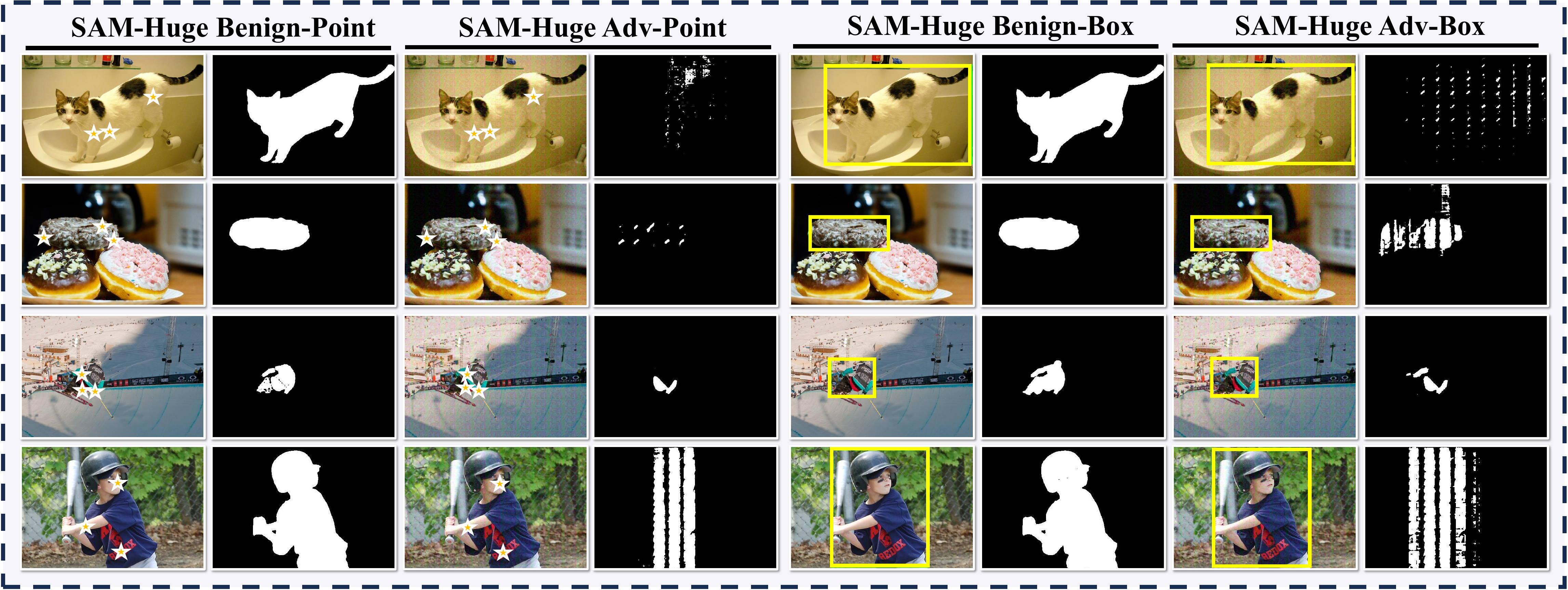}
\end{center}
    \caption{Visualization of DarkLLM for instruction-guided attacks on SAM-Huge.}
\label{fig:visualization_sam_h}
\end{figure*}

\begin{figure*}[!t]
\begin{center}
\includegraphics[width=0.98\linewidth]{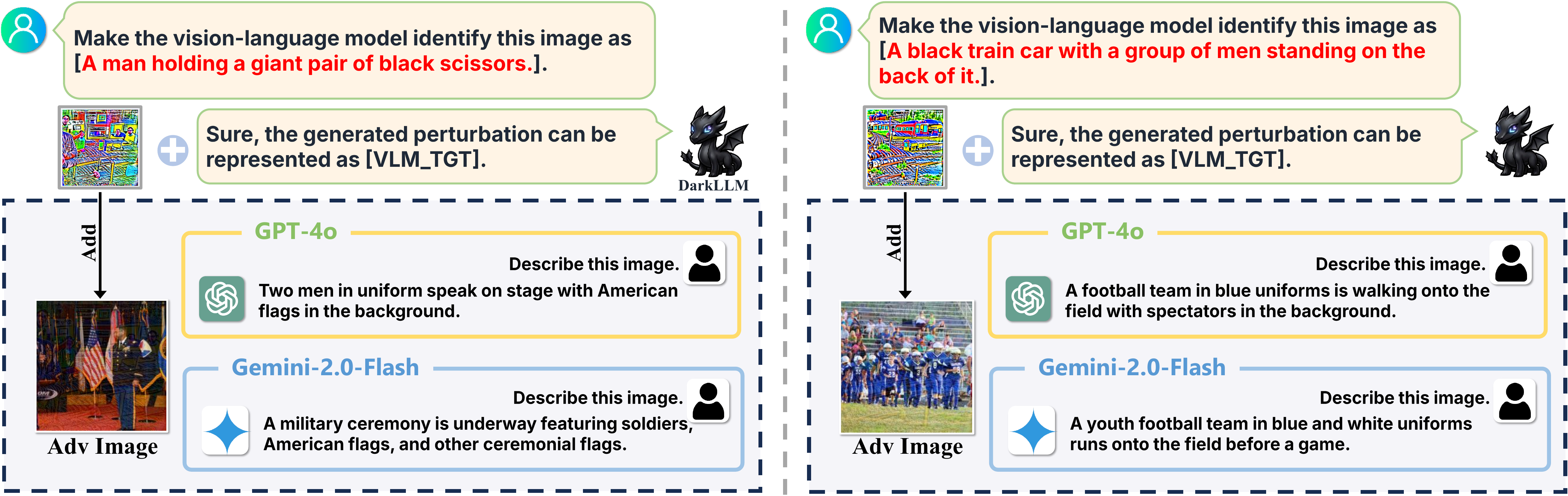}
\end{center}
    \caption{Failed cases of DarkLLM for instruction-guided attacks on Commercial MLLMs.}
\label{fig:visualization_failed_mllm}
\end{figure*}

\begin{figure*}[!t]
\begin{center}
\includegraphics[width=0.98\linewidth]{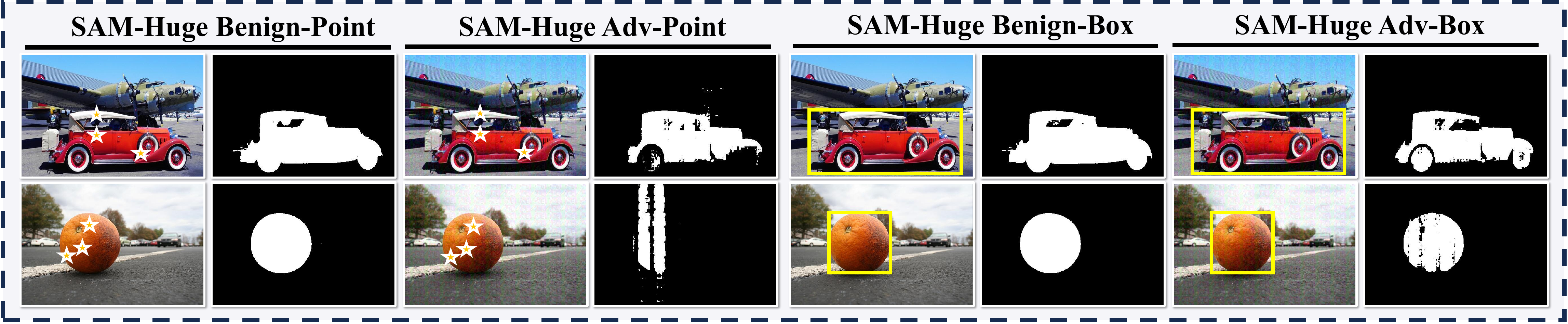}
\end{center}
    \caption{Failed cases of DarkLLM for instruction-guided attacks on SAM-Huge.}
\label{fig:visualization_failed_sam_h}
\end{figure*}

\begin{figure*}[h]
\begin{center}
\includegraphics[width=0.9\linewidth]{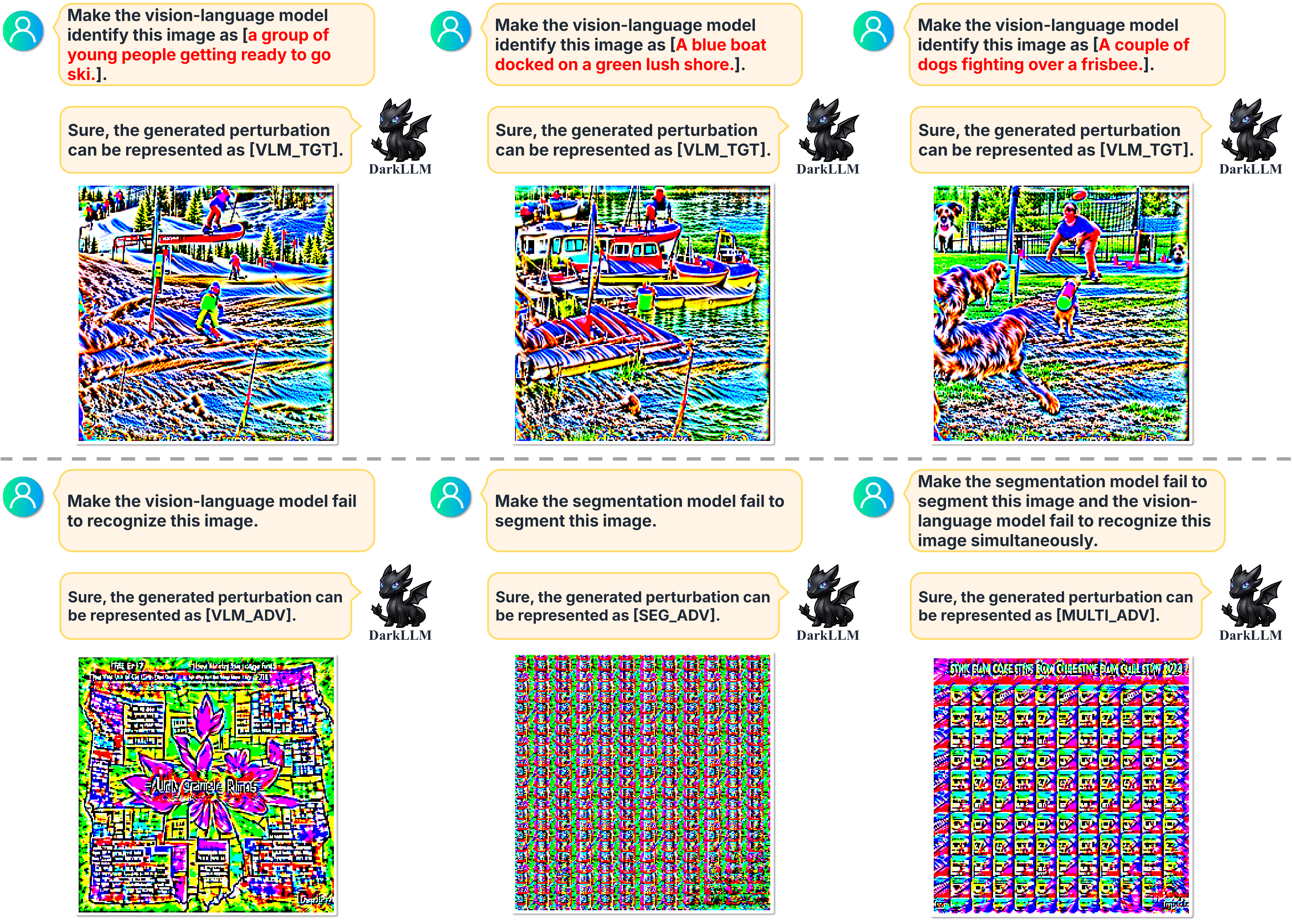}
\end{center}
    \caption{Details of perturbations generated by DarkLLM from user instructions.}
\label{fig:visualization_noise}
\end{figure*}

\section{Qualitative Analysis}

\noindent\textbf{Success Cases.}\quad
We provide further qualitative results to visually demonstrate the efficacy of DarkLLM. As shown in Figure~\ref{fig:visualization_mllm}, DarkLLM can generate diverse and potent targeted perturbations against frontier commercial MLLMs from simple natural language instructions, successfully forcing them to output the desired malicious content. Figures~\ref{fig:visualization_sam_b} to \ref{fig:visualization_sam_h} illustrate its impact on the SAM family. The adversarial noise completely neutralizes the segmentation capability of the surrogate model, SAM-Base, causing it to fail across various prompts. More importantly, this same perturbation exhibits excellent transferability, successfully misleading the larger and unseen SAM-Large and SAM-Huge backbones.

\noindent\textbf{Failure Cases.}\quad
Despite its strong overall performance, we also present failure cases to provide a transparent benchmark for future research. As shown in Figure~\ref{fig:visualization_failed_mllm}, the generated perturbation can be less effective on images with highly complex backgrounds; we hypothesize that the adversarial noise may be masked by the image's inherent high-frequency details. Similarly, Figure~\ref{fig:visualization_failed_sam_h} illustrates cases where a box prompt on a monochromatic object (an orange or a car) prevents the attack from successfully misleading the SAM-Huge model. We conjecture that the strong spatial prior provided by the box prompt, combined with the object's simple texture, makes the model more resilient to the perturbation in this specific scenario.
Nonetheless, as the first language-driven adversarial framework of its kind, we believe DarkLLM possesses substantial potential. These identified limitations do not detract from its core contribution but rather provide valuable insights that will inform the future development of both more sophisticated language-driven attacks and their corresponding defenses.

\clearpage
\newpage




\end{document}